\documentstyle[epsfig,epsf]{article}
\begin{document}
\begin{titlepage}
\begin{center}


{\large \bf Physics at the Interface of Particle Physics\\ and Cosmology }

\vskip 0.3in
R. D. Peccei\\
{\em Department of Physics and Astronomy, UCLA\\
 Los Angeles, CA 90095-1547    }

\end{center}
\date{}

\vskip .3in

\begin{abstract}
In these lectures I examine some of the principal issues in cosmology
from a particle physics point of view.  I begin with nucleosynthesis and
show how the primordial abundance of the light elements can help fix the
number of (light) neutrino species and determine the ratio $\eta$ of baryons
to photon in the universe now.  The value of $\eta$ obtained highlights
two of the big open problems of cosmology: the presence of dark matter and
the need for baryogenesis.  After discussing the distinction between hot
and cold dark matter, I examine the constraints on, and prospect for,
neutrinos as hot dark matter candidates.  I show next that supersymmetry
provides a variety of possibilities for dark matter, with neutralinos being
excellent candidates for cold dark matter and gravitinos, in some scenarios,
possibly providing some form of warm dark matter.  After discussing axions
as another cold dark matter candidate, I provide some perspectives on the
nature of dark matter before turning to baryogenesis.  Here I begin by
outlining the Sakharov conditions for baryogenesis before examining the
issues and challenges of producing a, large enough, baryon asymmetry at
the GUT scale.  I end my lectures by discussing the Kuzmin-Rubakov-Shaposhnikov
mechanism and issues associated with electroweak baryogenesis.  In particular,
I emphasize the implications that generating the baryon asymmetry at the
electroweak scale has for present-day particle physics.

\end{abstract}

\end{titlepage}

\section{Introduction}

There is a symbiotic relationship between particle physics and cosmology.
This is not surprising since both deal with physics in similar environments.
Cosmology, the physics of the early Universe, is concerned with matter at the
high temperatures characterizing the Universe at this epoch.  Particle
physics, the physics of fundamental constituents and their interactions,
deals with phenomena at very short distances which can only be probed at
high energy.  Since high temperatures and high energies are synonimous,
not surprisingly particle physics and cosmology are deeply intertwined.

If one begins to think of which aspects of particle physics are relevant
to the early Universe, one arrives soon at a very long list.  For convenience,
I have split up this list into four broad categories.  The first category
includes what might be called {\bf Planck scale physics}.  These are
interactions whose natural scale is of order of the Planck scale,\footnote{The
Planck mass $M_{\rm P}$ is the mass scale derived from Newton's constant.
In the natural system of unity we are using, where $\hbar=c=k=1$, $M_{\rm P}=
G_N^{-1/2}$.}
$M_{\rm P} \sim 10^{19}$ GeV, and which are of (likely) importance in the
early Universe.  A well known example is provided by Grand Unified Theories
(GUTs).  The second class revolves around the {\bf physics of light excitations}.  There are a variety of established, or postulated, nearly
massless particles like neutrinos and axions, which may well play an important
role in the Universe's energy density and could have had a role in creating
structure.  The third category encompasses {\bf stable, or long lived, heavy
particles}.  These particles, of which the LSP of supersymmetric theories is
a prime candidate, can be part of the constituents that make up the dark
matter of the Universe.  Or, if present in large enough quantities, as in 
the case of superheavy magnetic monopoles,
they can have a 
nefarious role in the evolution of the Universe. In the final category, I include the {\bf consequences
of symmetry breakdown}.  One suspects that phase transitions, of different
kinds (e.g. the one that gave rise to the inflationary phase of the
Universe), play a crucial role in the evolution of the Universe.  
In addition, the
breakdown of discrete symmetries, like CP, or of some continuous symmetries can
influence substantially the resulting cosmology.

Just as different aspects of particle physics affect the evolution of the
Universe, conversely the physics of the early Universe also has an important
bearing on particle physics.  That is, cosmological observations can help
inform particle theory.  For instance, as I will show below, the primordial
abundances of light elements effectively constrains the number of light
neutrino species.  Eventually, high precision data on the angular and
power spectrum of the cosmic microwave background radiation should help pin
down the neutrino mass spectrum.  Similarly, as we will see, the precise
nature of baryogenesis deeply influences the view one has of the sources for
CP violation.

In many instances, the back and forth relation between particle physics and
cosmology has proven very stimulating for both fields.  Baryogenesis provides
perhaps the best example of this symbiotic relationship.  The Sakharov
conditions for baryogenesis in the Universe, enunciated in
1967\cite{Sakharov}, were first made manifest in GUTs about a decade later
and contributed to the enormous interest in these theories.  However, GUTs
also overproduced magnetic monopoles\cite{Preskill} creating a cosmological
crisis which was only resolved through the development of the inflationary
Universe scenario\cite{Guth}.  Although it was pointed out by 
't Hooft\cite{'tHooft} already in 1976 that, as a result of the chiral
anomaly, baryon number is not exactly conserved in the Standard Model, the
rate for these processes seemed insignificantly small to be much more than a
curiosity.  However, about a decade later Kuzmin, Rubakov and
Shaposhnikov\cite{KRS} showed that these processes could be important at
temperatures near the electroweak phase transition, opening up the possibility
that baryogenesis occurred much later in the evolution of the Universe than
hereto believed.  Bounds on the Higgs mass obtained at LEP in the 1990's,
however, suggested that this interesting cosmological scenario was only tenable
if there were additional CP violating phases, besides the usual CKM phase of
the Standard Model.  What the next development in this saga will be is
unclear.  Nevertheless, it is obvious that, at least in this area, cosmology
and particle physics are deeply intertwined.

\section{Primordial Nucleosynthesis and the Number of Neutrino Species}

I begin my lectures by discussing nucleosynthesis.  Although this material
is well known\cite{KT}, its affords me a way to introduce, in a familiar
context, a number of concepts which will be of use later.  Furthermore,
nucleosynthesis is also the first area where a cosmological observation
had a direct bearing on particle physics\cite{Schramm}, so it makes sense to
begin here.

One has known for a fairly long time that the bulk of the Helium present in the
Universe is primordial\cite{Helium}.  Although a small amount of the
approximately 25\% mass fraction of Helium was generated in stars, all the
rest was generated by nucleosynthesis in the early Universe.  The calculation
of this primordial fraction of Helium, $Y_{\rm P}$, by 
Wagoner, Fowler, and Hoyle\cite{FHW} 
in the late 60's was one of the early triumphs of cosmology
and remains an important milestone for our understanding of the Universe.
When one examines the ingredients that lead to a prediction of
$Y_{\rm P} \sim 0.25$, two play a crucial role.  These are the energy density
of the Universe at the time when the neutrons and protons go out of
equilibrium and the temperature where enough deuterium is formed.  The former,
in detail depends on the number of light neutrino species $N_\nu$.  The
latter is related to $\eta = n_{\rm B}/n_\gamma$, the ratio of baryons to
photons in the Universe now.  As we shall see, the ratio $\eta$ is an
important cosmological parameter, related both to the quantity of dark matter
in the Universe and to the asymmetry between matter and antimatter in the
Universe.  $N_\nu$, on the other hand, is
a crucial number for particle physics.  This quantity
is now known to great accuracy as a result of precise measurements of the
width of the $Z$ boson.  However, before these measurements $N_\nu$ 
already could be
determined reasonably well indirectly through its numerical influence in
predicting the Helium mass fraction $Y_{\rm P}$\cite{Schramm}.

I will sketch now the calculation of $Y_{\rm P}$, focusing particularly on
how the final answer depends on $N_\nu$ and $\eta$.  The crucial concept to
understand is the idea of {\bf freeze-out}, or decoupling, of physical
processes in the evolution of the Universe\cite{KT}.  This occurs when the
interaction rate $\Gamma = n\langle\sigma v\rangle$ for the process in
question becomes slower than the Universe's expansion rate.  In the standard
Big Bang cosmology, this latter rate is given by the Hubble parameter $H$,
which scales with the Universe's temperature as
\begin{equation}
H\sim \frac{T^2}{M_{\rm P}}
\end{equation}
If $\Gamma$ for certain processes is much 
greater than $H$, then these processes
are in equilibrium in the Universe.  Conversely, if $\Gamma \ll H$, the
interaction rate is too slow compared to the Universe's expansion rate to keep these processes in equilibrium.  The freeze-out temperature is the temperature
at which $\Gamma \simeq H$.  That is, it is the temperature (or time) when
certain processes begin to go out of equilibrium in the Universe.

There are two important moments for nucleosynthesis.  The first of these is
related to the freeze-out of the weak interactions between neutrons and
protons.  Above this freeze-out temperature, neutrons and protons are
in equilibrium through the weak interactions
\begin{eqnarray}
n&+&\nu_e \leftrightarrow p+e \nonumber \\
n&+&e^+ \leftrightarrow p+\bar\nu_e \nonumber \\
n &\leftrightarrow& p+e+\bar\nu_e~.
\end{eqnarray}
The rate for these processes scales as 
$\Gamma\sim G_FT^5$, and the ratio of neutrons to protons is fixed by
their mass difference, through the usual Boltzmann factor
\begin{equation}
n/p = e^{-\Delta m/T}~.
\end{equation}
Freeze-out occurs when the Universe cools to a temperature of around
$10^{10}~^\circ{\rm K} \simeq 1$ MeV\cite{KT}, when $\Gamma\simeq H$.
The freeze-out temperature $T^*$ fixes the ratio of neutron to baryons at
that time in terms of the Boltzmann factor:
\begin{equation}
X_n(T^*) \equiv \frac{n}{n+p}\bigg|_{T^*} =
\frac{e^{-\Delta m/T^*}}{1+e^{-\Delta m/T^*}} \simeq 0.23~.
\end{equation}

The Helium mass fraction $Y_{\rm P}$ depends on $X_n(T^*)$, and the
particular value of $X_n(T^*)$ one obtains depends in detail on
$N_\nu$.\footnote{The value given in Eq. (4) corresponds to that obtained for
$N_\nu=3$.}  
This latter assertion is easily verified by examining Einstein's
equations for a Friedmann-Robertson Walker Universe,\footnote{In Eq. (5),
$R$ is the scale parameter characterizing the FRW Universe.\cite{KT}
The curvature
term, at this early stage of the Universe, can be safely neglected and is
omitted from this equation.}
which relate the Hubble parameter to the matter density:
\begin{equation}
H^2 = \left(\frac{\dot R}{R}\right)^2 =
\frac{8\pi G_N}{3} \rho~.
\end{equation}
The matter that drives the expansion of the Universe at this time 
is composed of the
states which are still relativistic then: the photons, the electrons and
positrons, and the $N_\nu$ species of neutrinos.  Thus
\begin{equation}
\rho = \rho_\gamma + \rho_{e^\pm} + \rho_{\nu/\bar\nu} =
aT^4\left\{1 + \frac{7}{4} + \frac{7}{8} N_\nu\right\} =
aT^4\frac{(22+7N_\nu)}{8}~.
\end{equation}
Here $a$ is the Stefan-Boltzmann constant and the different weights take
into account the different statistics between bosons and fermions and the
fact that neutrinos have only one active helicity component.\cite{KT}  From (5) and
(6) one sees that the expansion rate $H$ at $T^*$ scales as 
$H\sim (22+7N_\nu)^{1/2}T^{*2}/M_{\rm P}$.  Since $\Gamma \sim G_FT^{*5}$,
it follows that $T^*$ depends on $N_\nu$ as
\begin{equation}
T^* \sim (22+7N_\nu)^{1/6}~.
\end{equation}
In view of the above and Eq. (4),
one sees that the neutron to baryon ratio at
freeze-out $X_n(T^*)$ increases if the number of neutrino species $N_\nu$
increases.

After neutron-proton freeze-out, the ratio $X_n$ decreases exponentially
because of neutron decay, so that at any time $t$ after $t^*$
\begin{equation}
X_n(t) = X_n(t^*) e^{-\frac{(t-t^*)}{\tau_n}}~,
\end{equation}
with $\tau_n$ being the neutron lifetime.
Helium nucleosynthesis occurs at a time $t_d$ (or temperature $T_d$) when
enough deuterium is formed ($X_d \sim X_n$), since (almost) all
deuterium transmutes directly into Helium through the reaction
$d+d \to {\rm He} + \gamma$.  Because the reaction $n+p\leftrightarrow d+
\gamma$ has a fast rate, the deuterium fraction is fixed by a Boltzmann factor.
One has
\begin{equation}
\frac{X_d}{X_nX_p}\bigg|_T = n_{\rm B}(T) \frac{3}{2\sqrt{2}}
\left(\frac{2\pi}{M_NT}\right)^{3/2}
 e^{{\rm B}/T}~.
\end{equation}
In the above B is the deuterium binding energy, ${\rm B}\simeq 2.2$ MeV,
and $n_{\rm B}(T)$ is the density of baryons at the temperature $T$.
Nucleosynthesis starts at $T = T_d$ when the ratio above is of $O(1)$.
Thus,
the temperature $T_d$ is intimately related to the baryon number density
at that stage of the Universe.  In turn, one can relate
$n_{\rm B}(T_d)$ to the density of photons at that epoch $n_\gamma(T_d)$
which just depends on $T_d$, $n_\gamma(T_d) \sim T_d^3$. 
The argument is simple.  Because the ratio
of the baryon to photon 
densities is {\bf independent} of temperature, knowing the
baryon to photon ratio now, $\eta$, it follows that
\begin{equation}
n_{\rm B}(T_d)=\eta n_\gamma(T_d) \sim \eta T_d^3~.
\end{equation}

Numerically, one finds\cite{KT} that $T_d\sim 10^9~{^\circ{\rm K}} \sim
0.1$ MeV.  One sees from Eqs. (9) and (10) that if the ratio $\eta$ decreases,
so does the temperature $T_d$ when nucleosynthesis starts.  Basically, for
smaller $\eta$ one needs a larger Boltzmann factor, $e^{{\rm B}/T_d}$.
At $T_d$, because of neutron decays, the ratio $X_n(T_d)\simeq 0.12$,
roughly half of what it was at freeze-out.  Since at $T_d$,~
$X_n\sim X_d$ and all the deuterium is transmuted into Helium, one expects
\begin{equation}
Y_{\rm P} \simeq 2X_n(T_d) \sim 0.24~.
\end{equation}
\begin{figure}
\centerline{
\epsfig{file=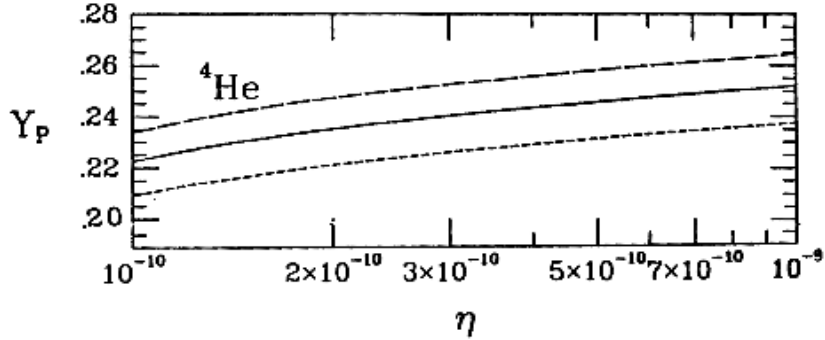,width=5in}
}
\caption[]{Predicted values for $Y_P$ for different values of $N_{\nu}$}
\end{figure}

Precise results for the Helium mass fraction $Y_{\rm P}$ depend on the
actual value of $N_\nu$ and $\eta$ (as well as on the neutron lifetime,
$\tau_n$).  As we saw, $X_n(T^*)$ is larger the larger $N_\nu$ is.  Thus
$Y_{\rm P}$ increases with increasing $N_\nu$.  Similarly, a larger $\eta$
also leads to a larger $X_n(T_d)$ and hence a larger $Y_{\rm P}$.  Fig.
1 shows the results of a detailed calculation \cite{Chic} of the
primordial abundance of He, plotted as a function of $\eta$, for three
different values of $N_\nu$ ($N_\nu = 2,3,4$).  Clearly, if one knew
$\eta$, from estimates of $Y_{\rm P}$ one could infer a value for $N_\nu$.
One way to infer $\eta$ is to compute also the primordial abundances of
other light elements, besides Helium, 
produced by nucleosynthesis.  Demanding
concordance of these results fixes $\eta$ and one can then infer a value for
$N_\nu$ from cosmology\cite{Schramm}.  The results of the Chicago 
group\cite{Chicago} in the early 1980's, shown in Fig. 2, using a range
$0.22 < Y_{\rm P} < 0.26$ for the Helium abundance, predicted
\begin{equation}
N_\nu \leq 4~.
\end{equation}

\begin{figure}
\centerline{
\epsfig{file=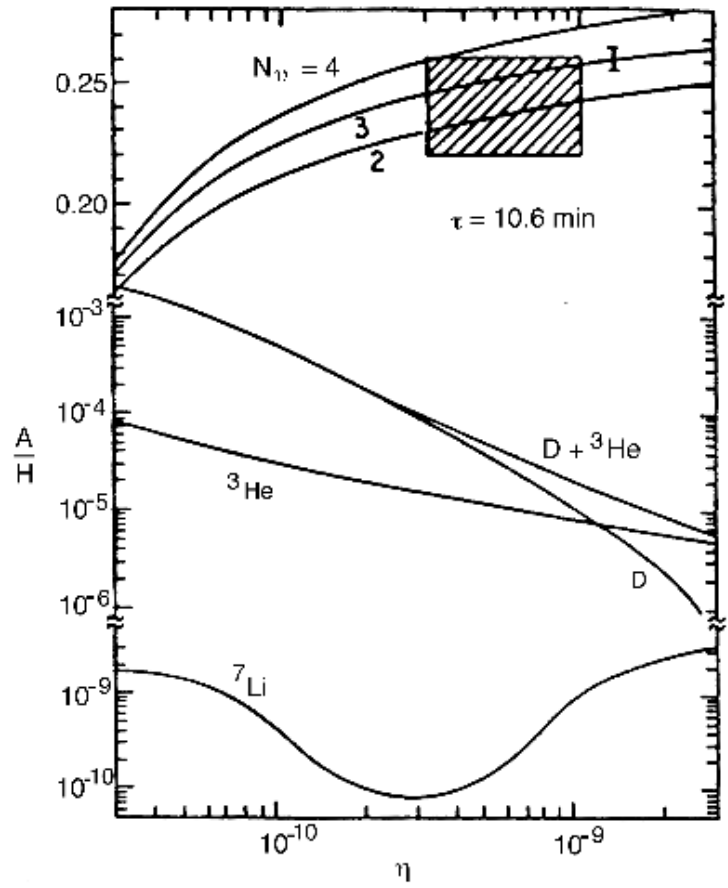,width=5in}
}
\caption[]{Primordial abundance of light elements as a function of $\eta$}
\end{figure}

This cosmological inference was verified at LEP and the SLC almost a decade
later, by studying $e^+e^-$ scattering at the energy of the $Z$ boson
mass.  The amplitude for the annihilation of $e^+$ and $e^-$ into a 
fermion-antifermion pair depends on the width of the $Z$
\begin{equation}
A(e^+e^-\to f\bar f) \sim
\frac{1}{s-M_Z^2+i\Gamma_ZM_Z}~.
\end{equation}
This width, in turn, depends on the number of light neutrinos species---where
light here means $m_\nu \ll \frac{M_Z}{2}$.  One has:
\begin{equation}
\Gamma_Z = \Gamma(Z\to \hbox{charged states}) + N_\nu \Gamma(Z\to 
\nu\bar\nu)~.
\end{equation}
Fig. 3 plots some early data from the ALEPH\cite{Aleph} collaboration at 
LEP which clearly shows that $N_\nu = 3$ is preferred.  The most recent
compilation of results from all the four LEP collaborations\cite{electroweak}
doing a Standard Model fit, gives
\begin{equation}
N_\nu = 2.993\pm 0.011~.
\end{equation}
A less accurate, but more direct measurement of the, so-called, invisible
width of the $Z$---assuming that this width
is due to $Z$ decays into neutrino pairs---gives
instead
\begin{equation}
N_\nu = 3.09\pm 0.13~.
\end{equation}
Thus, there is now strong evidence that $N_\nu = 3$.

\begin{figure}
\centerline{
\epsfig{file=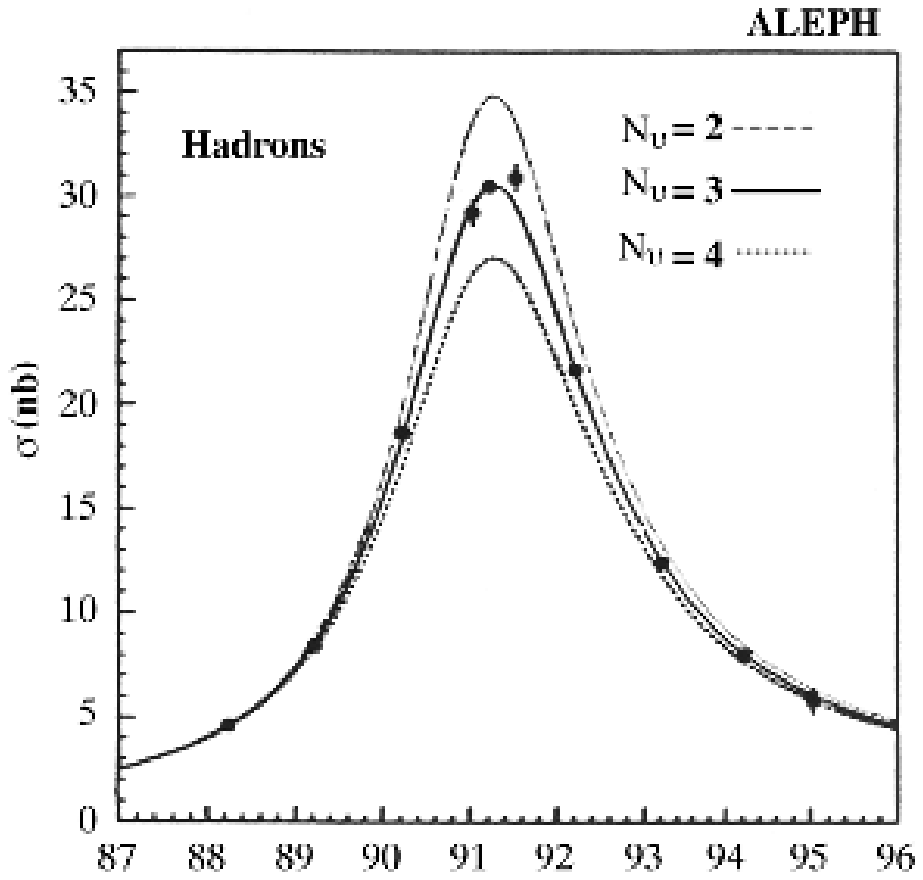,width=5in}
}
\caption[]{Plot of $\sigma(e^+e^-\rightarrow {\rm hadrons})$ at the Z-resonance}
\end{figure}

Given these results, the present-day discussions of nucleosynthesis take
$N_\nu = 3$ and try to get stronger limits on the baryon to photon ratio
$\eta$ from the demand of concordance of all the primordial abundances.
A recent example of such an analysis is the work of Copi, Schramm and
Turner\cite{CST}, whose results are depicted in Fig. 4, yielding for
$\eta$ the range
\begin{equation}
2.4\times 10^{-10} < \eta \leq 4.2\times 10^{-10}~.
\end{equation}

\begin{figure}
\centerline{
\epsfig{file=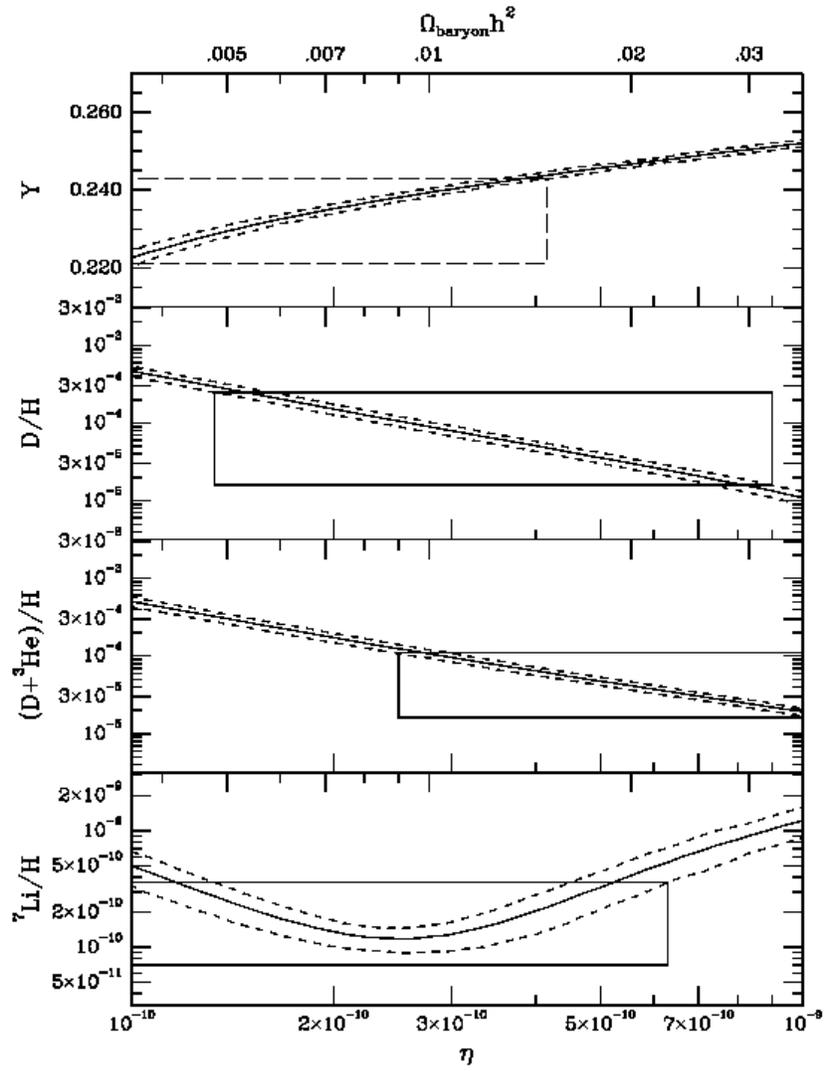,width=5in}
}
\caption[]{Concordance of meausured primordial element abundances, from \cite{CST}}
\end{figure}

A recent study of updated data for primordial $^4{\rm He}$ by
Olive, Skillman and Steigman\cite{OSS} pins down $Y_{\rm P}$ in a narrow
range
\begin{equation}
Y_{\rm P} = 0.234\pm 0.002\pm 0.005~,
\end{equation}
where the first error is statistical and the second is an estimate of the
possible systematic error.  Using the above, the 95\% confidence limit for
$Y_{\rm P}$ gives $Y_{\rm P} < 0.244$, which allows Olive, Skillman and
Steigman\cite{OSS} to set a 95\% CL for $\eta$ of
\begin{equation}
\eta < 3.8\times 10^{-10} ~~~~~~~~~(95\% ~{\rm CL})~.
\end{equation}
This result is in agreement with the recent work on the primordial
abundance of deuterium obtained by studying quasi-stellar objects
(QSO) absorption lines\cite{HighD}, which infers a rather high
primordial deuterium abundance.  However, very recent work by Tytler
{\it et al.}\cite{Tytler}, based on two correlated QSO observations, obtains
a discordant, very low, primordial deterium abundance yielding large $\eta$
values:
\begin{equation}
5.1\times 10^{-10} < \eta < 8.2\times 10^{-10}~.
\end{equation}
Such values
correspond to a range for $Y_{\rm P}$ ($0.246 < Y_{\rm P} < 0.282$)
above the 95\% limit of Olive, Skillman and Steigman\cite{OSS}.

Clearly, the situation at the moment is still unsettled and it is difficult
to draw strong inferences.  Possibly, the simplest assumption to make is
that the actual value for $Y_{\rm P}$ is subject to much stronger systematic
uncertainties that those assumed by Olive, Skillman and Steigman\cite{OSS}.
In what follows, we shall take the value obtained by Copi, Schramm and 
Turner\cite{CST} for $\eta$ but, following their suggestion, shall boost it to
cover a $2\sigma$ range.  Then one has
\begin{equation}
1.9\times 10^{-10} < \eta < 5.8 \times 10^{-10} ~~~~~~~(2\sigma ~\hbox{range})~,
\end{equation}
which is a range broad enough to encompass all recent determinations.

\section{Two Open Problems in Cosmology: Dark Matter and Baryogenesis}

The ratio $\eta$, which we just saw is important for nucleosynthesis, lies
at the heart of two of the biggest open problems in cosmology today,
those of dark matter and of baryogenesis.  Recall that $\eta$ was the ratio
of the number density of baryons to photons in the Universe now.  The photon
density itself is extremely well known from the measurement of the
temperature of the cosmic background radiation\cite{CBR}
\begin{equation}
T_\gamma = (2.726\pm 0.005)^\circ{\rm K}~,
\end{equation}
yielding
\begin{equation}
n_\gamma = \int \frac{q^2dq}{\pi^2}
\frac{1}{(e^{q/T_\gamma}-1)} = [0.625~T_\gamma]^3 \simeq
400~{\rm cm}^{-3}~.
\end{equation}
Therefore a value for $\eta$ serves to fix the energy density of baryons
in the Universe now:
\begin{equation}
\rho_{\rm B} = m_Nn_{\rm B} = m_N\eta n_\gamma~,
\end{equation}
where $m_N$ is the nucleon mass.  Using Eqs. (21) and (23) one finds
\begin{equation}
1.3\times 10^{-31}~{\rm g/cm}^3 < \rho_{\rm B} <
4\times 10^{-31}~{\rm g/cm}^3~.
\end{equation}
This value is interesting since, as we shall see below, it is a few
percent of what is needed to close the Universe.

If one does not neglect the curvature term, Einstein's equations in a
Friedmann Robertson Walker Universe have the form
\begin{equation}
H^2 = \left(\frac{\dot R}{R}\right)^2 =
\frac{8\pi G_N}{3} \rho - \frac{k}{R^2}~.
\end{equation}
The constant k here describes the geometry of the Universe.  If $k > 0$,
the Universe is closed.  If $k < 0$, it is open.  Finally, if $k$ vanishes,
one has a Universe with no curvature---a flat Universe.  It is useful to
consider the quantity $\Omega(t)$, which is essentially the ratio of the
matter density to the square of the Hubble parameter
\begin{equation}
\Omega(t) = \frac{\rho}{\left[\frac{3}{8\pi}\frac{H^2}{G_N}\right]}~.
\end{equation}
Using Einstein's equations, one sees that $\Omega(t)$ 
characterizes the geometry
\begin{equation}
\Omega(t) = \frac{1}{1-X(t)}~; ~~~~~
X(t) = \frac{\left[\frac{k}{R^2}\right]}
{\left[\frac{8\pi G_N}{3}\rho\right]}~,
\end{equation}
with $\Omega > 1$ corresponding to a closed Universe, and $\Omega < 1$
corresponding to an open Universe.
The value of $\Omega(t)$ at the present time $\Omega_o$ depends on how
$\rho$ compares to the, so-called, critical density
\begin{equation}
\rho_o = \frac{3H_o^2}{8\pi G_N}~,
\end{equation}
with $H_o$ being the value of the Hubble parameter now---the Hubble
constant.

From Eq. (28) one sees that a flat
Universe has $\Omega(t)=1$.  Thus the ratio of the Universe's
density now to the critical density:
\begin{equation}
\Omega_o = \frac{\rho}{\rho_o}
\end{equation}
directly informs one about the Universe's geometry,
with $\Omega_0=1$ corresponding to a flat Universe.  Unfortunately, the
Hubble constant itself is not that easily determined.  Conventionally,
one writes:\footnote{A Mega parsec (Mpc) is $3\times 10^6$ light years.}
\begin{equation}
H_o = 100~{\rm h} \frac{{\rm Km}}{\rm{Mpc~sec}}~,
\end{equation}
and one typifies the uncertainty in $H_o$ through a range for h.
Traditionally, this uncertainty
corresponds to h lying in the range 
\begin{equation}
0.5 < {\rm h} < 1~,
\end{equation}
although a more modern determination\cite{Hubble} gives
\begin{equation}
h = 0.6 \pm 0.1~.
\end{equation}
Numerically, one finds that the critical density has the value
\begin{equation}
\rho_o = 1.9\times 10^{-29}~{\rm h}^2~{\rm g/cm}^3~.
\end{equation}
If the density of the Universe is above $\rho_o$ the Universe is closed.
Clearly, if baryons dominate the energy density of the Universe, Eq. (25)
tells us that the Universe is open.

If we denote the baryonic contribution to $\Omega_o$ by
$\Omega_{\rm B} = \rho_{\rm B}/\rho_o$, using Eq. (25) and (34), one 
has\cite{CST}
\begin{equation}
0.007 \leq \Omega_{\rm B}{\rm h}^2 \leq 0.021~.
\end{equation}
This equation is remarkable in several ways.  First, it appears that
$\Omega_{\rm B}$ itself is much bigger than the value one would infer from
the amount of {\bf luminous matter} in the Universe.  Using
h = $0.6\pm 0.1$, from Eq. (35) one sees that $\Omega_{\rm B}$ ranges from
about 0.014 to 0.084.  On the other hand, the best estimates of the fraction 
of luminous matter in the Universe\cite{luminous} give a range
\begin{equation}
\Omega_{\rm luminous} \simeq 0.003-0.017~,
\end{equation}
half an order of magnitude smaller.  So one infers that there is
substantial non-luminous baryonic dark matter.  The existence of this dark
matter is also inferred from the observed flat rotation curves in spiral
galaxies.  Normally, outside the luminous body of the galaxy, one would
expect the circular velocity to drop as $r^{-1/2}$, but it does not, as shown
in Fig. 5.  From these measurements, one deduces values\cite{rotation}
\begin{equation}
\Omega_{\rm rot.~curves} \simeq 0.03-0.10~,
\end{equation}
much more comparable to those for $\Omega_{\rm B}$.

\begin{figure}
\centerline{
\epsfig{file=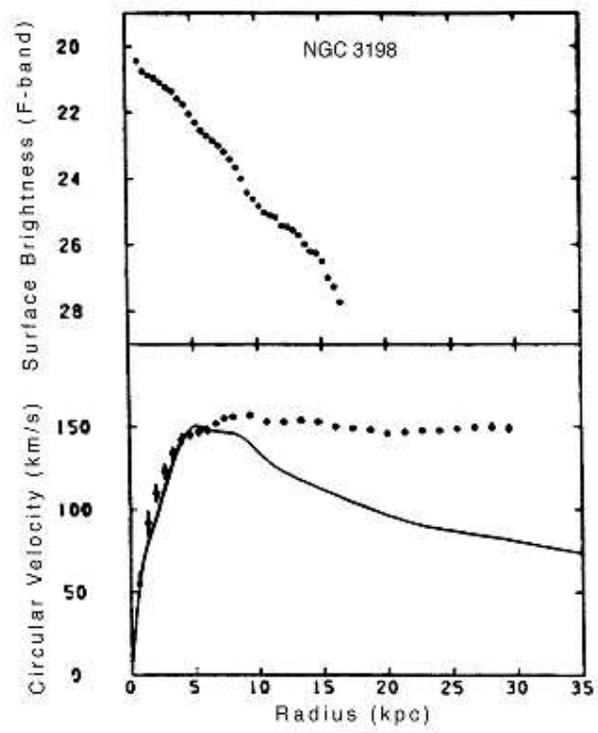,width=4in}
}
\caption[]{A typical flat rotation curve extending beyond the luminous body of the galaxy}
\end{figure}

Second, since $\Omega_{\rm B}$ is much bigger than the contribution to the
energy density made by photons, neutrinos and electrons,\footnote{The energy
density of photons and neutrinos follows directly from the temperature
of the cosmic microwave background radiation $T_\gamma$, with
$\rho_\gamma = a T_\gamma^4$ and $T_\nu = (4/11)^{1/3}T_\gamma$ due to photon
reheating.\cite{KT}  Charge neutrality requires $n_e = n_{\rm p}$ and
hence, because of the proton-electron  mass difference, the energy density
of electrons in the Universe is negligible.} if $\Omega_o\simeq \Omega_{\rm B}$
then there is an enormous fine-tuning problem.  
In this case, the parameter $X(t)$ now, $X_o$,
is roughly of order ten to one hundred:
\begin{equation}
X_o = 1-\frac{1}{\Omega_o} \simeq -\frac{1}{\Omega_{\rm B}}~.
\end{equation}
However, $X(t)$ being the ratio of the curvature term to the energy density
term [cf. Eq. (28)] scales as\footnote{I neglect here, for simplicity,
the fact that matter dominates over radiation in the latter stages of the
Universe.  This changes the fine-tuning problem only qualitatively, not
quantitatively.} $X(t)\sim R^2(t)\sim T^{-2}$.  Hence $X(t) \simeq
-1/\Omega_{\rm B}~(T_o/T)^2$ and therefore
\begin{equation}
\Omega(T) \simeq \frac{1}{1+\frac{1}{\Omega_{\rm B}}
\left(\frac{T_o}{T}\right)^2}~.
\end{equation}
To get $\Omega_o \simeq \Omega_{\rm B}$ now, in the early Universe the density
must have been unbelievably close to the critical density.  For instance
at the Planck temperature, $T_{\rm P} \simeq 10^{32}~^\circ{\rm K}$,
$\Omega(T_{\rm P}) \simeq 1-~ O(10^{-62})$~!

The solution to the fine-tuning problem above is provided by inflation.\cite{Guth}  In an inflationary Universe, there is an 
exponential 
growth of the scale factor at early times.  Effectively then the
curvature term ${\rm k}/R^2$ is totally negligible and the latter evolution
of the Universe corresponds to that of a flat Universe, with
${\rm k}_{\rm eff} = 0$.  Thus, if one wants to avoid fine-tuning
as a result of inflation, then $\Omega(t) = 1$ and the Universe is
always at the critical density.  In this case, the value of $\Omega_{\rm B}$
obtained, since it is in the percent range, tells us that the Universe is
dominated by {\bf non-baryonic dark matter}.

\begin{figure}
\centerline{
\epsfig{file=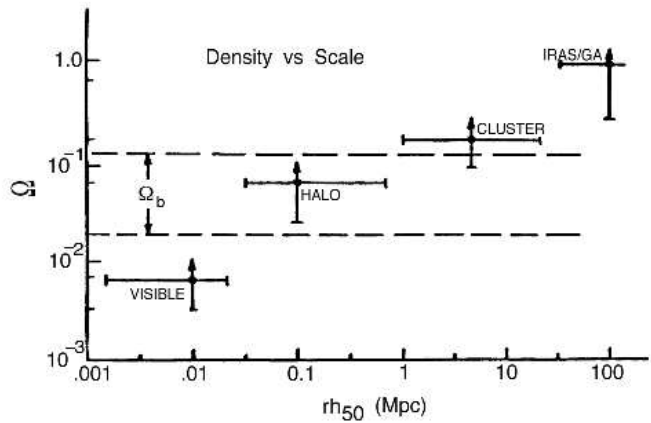,width=4in}
}
\caption[]{Variation in $\Omega_0$ as a function of the scale of the structures meausured}
\end{figure}

Besides the theoretical bias for considering $\Omega = 1$, there is
actually observational evidence for $\Omega$ being greater than
$\Omega_{\rm B}$, obtained by reconstructing the energy density from the
flow of peculiar velocities in superclusters of galaxies.  It appears that
the values of $\Omega_o$ one infers are largest when one measures 
the density on the
largest structures
in the Universe, as shown in Fig. 6.  All the data on $\Omega_o$
has been summarized recently by Dekel, Burnstein and White\cite{BDW} who, if
one assumes that there is no cosmological constant, give the following
range for this quantity:
\begin{equation}
0.3 \leq \Omega_o \leq 1.3~.
\end{equation}
The lower bound above comes from the cosmic velocity flows, while the upper
bound comes from the age of the Universe (assuming h = $0.6\pm 0.1$).  Fig.
7 summarizes these results, allowing for the possibility of a cosmological
constant.

\begin{figure}
\centerline{
\epsfig{file=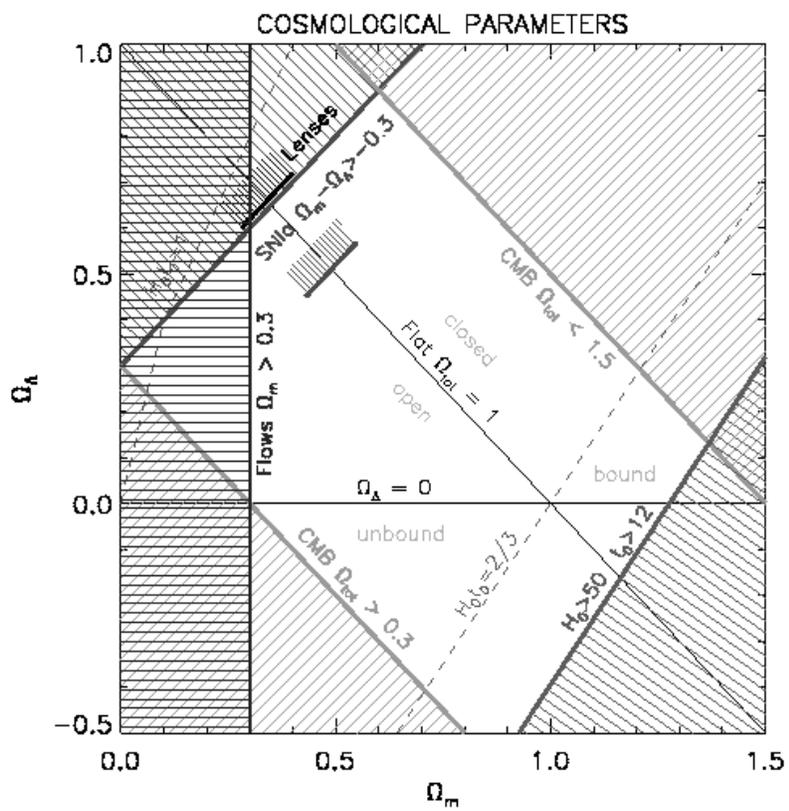,width=5in}
}
\caption[]{Summary of present status of $\Omega_0$, from \cite{BDW}}
\end{figure}

Very recent data on type-I supernovas at high redshift \cite{SNdata} \cite{Perl} has provided some evidence,  at the 3$\sigma$ level, that the Universe's expansion is actually {\bf{ accelerating}} rather than decelerating. Since the deceleration parameter\cite {weincosm} meausures the difference between the contribution of matter to $\Omega$ and that of the cosmological constant
\begin{equation}
q^0 = \frac{\Omega_{M}}{2}-\Omega_{\Lambda},
\end{equation}
this data suggests a non vanishing cosmological constant. However, these results are based on the assumption that supernovas are standard candles. Further, they are critically dependent on the highest redshift supernovas observed. There is also some disagreement among the results of the two groups. For a flat Universe, \cite{SNdata} obtains 
$\Omega_{M}= 0.32\pm 0.1$ and $\Omega_{\Lambda}=0.68 \pm 0.1$, while \cite{Perl} find
$\Omega_{M}= 0.6\pm 0.2$ and $\Omega_{\Lambda}=0.4 \pm 0.2$. In my view, it is probably to early to abandon the idea of a Universe where only matter  (of all types) contributes to give $\Omega=1$. However, these results give one pause.

The parameter $\eta$, besides fixing $\Omega_{\rm B}$ and adumbrating
the dark matter problem, has another role.  $\eta$ is also a measure of the
amount of matter-antimatter asymmetry in the Universe.  From observation, it
appears that the Universe is matter dominated, with little or no antimatter.\cite{CDRG}
The observed antiprotons in cosmic rays, whose typical ratio to protons is of
$O(\bar p/p\sim 10^{-4})$, are entirely consistent with the flux coming from
pair production.  Furthermore, no characteristic $\gamma$-rays are seen in
the sky which could arise from $p-\bar p$ annihilations.  If the Universe had
islands of antimatter, one would expect such signals to be present.  In addition,
there are also theoretical difficulties in assuming
that the Universe was matter-antimatter symmetric in its late evolution.
In this case, one can estimate the amount of matter that would remain after
the $p$ and $\bar p$ in the Universe go out of equilibrium around
$T \sim O(1~{\rm GeV})$.  Below this temperature inverse annihilations
($2\gamma\to p+\bar p$) are blocked and the direct process
$p+\bar p\to 2\gamma$ considerably reduces the number of protons (and
antiprotons) compared to that of photons to values of $\eta\sim 10^{-18}$\cite{KT}.  

For these reasons, $\eta\sim O(10^{-10})$, as observed, is
evidence that there was some {\bf primordial baryon asymmetry}.  That is,
really,
\begin{equation}
\eta = \frac{(n_{\rm B}-n_{\bar{\rm B}})}{n_\gamma}~.
\end{equation}
If the value of $\eta$ were to codify an initial asymmetry for the Universe,
this would appear to be a pretty mysterious initial condition.  Fortunately,
as Sakharov\cite{Sakharov} first pointed out, it is possible to generate such
a baryon-antibaryon asymmetry dynamically and so $\eta$ can be a reflection
of some primordial processes.  To generate such an asymmetry dynamically, as we
will discuss later on in much greater detail, the underlying theory must
violate baryon number, as well as C and CP.  Thus it appears that even though
$\eta$ is an important cosmological parameter, its origins are tied to
particle physics also!

\section{Hot and Cold Dark Matter}

An important classification scheme for dark matter is whether the relic
dark matter candidates were created by a thermal process or as a result
of some non-thermal process (e.g. in a phase transition).  Thermal relics
can be further distinguished by whether they were relativistic or
non-relativistic at the time their interaction rate fell below the
Universe's expansion rate.  Relics which were relativistic at freeze-out
are labeled hot dark matter (HDM), while relics which were non-relativistic
at freeze-out are called cold dark matter (CDM).\footnote{There is also
warm dark matter (WDM).  These are relics which, while relativistic at
freeze-out, have much weaker interaction rates and so, in some sense, have
also some of the characteristics of CDM.}

Particle physics provides possible dark matter candidates in all these
categories.  Neutrinos, neutralinos and gravitinos are thermal relics,
while axions are an example of a non-thermal relic.  Neutrinos are a 
prototypical hot dark matter relic.  Neutralinos are an example of
cold dark matter, while gravitinos are warm dark matter candidates.  Because
only zero momentum axions can contribute substantially to the Universe's
energy density, axions are also cold dark matter candidates.  In what
follows, I will describe some of the characteristics of these possible
dark matter candidates.

The contribution of thermal relics to $\Omega$ depends on what their abundance
was when their interaction rate fell below the Universe's expansion rate.
Freeze-out occurs when, for the relic $\chi$, $\Gamma_\chi \simeq H$.  For
{\bf hot relics} the freeze-out temperature is much greater than the mass of
the relic:  $T_\chi \gg m_\chi$.  In this case, at freeze-out 
$n_\chi\sim n_\gamma$.  Because the density of the relic to that of photons
is an invariant, the contribution to $\Omega_0$ of any hot dark matter
relic is just a function of its mass (and $T_\gamma^0 \simeq 3^\circ{\rm K}$). Calling this contribution $\Omega_{\chi}$, one finds
\begin{equation}
\Omega_\chi[{\rm HDM}] = \frac{m_\chi n_\chi^0}{\rho_0} \simeq
\frac{m_\chi n_\gamma^0}{g^*\rho_0} \simeq
\frac{1}{g^*}\left[\frac{m_\chi}{92~{\rm eV}}\right]
\frac{1}{h^2}~.
\end{equation}
Here $g^*$ counts the effective degrees of freedom at freeze-out, with $g^*=1$
for neutrinos.  The above shows that particles with eV masses
can be cosmologically significant, contributing substantially to the
Universe's energy density.  This observation was first made about 25 years
ago by Cowsik and McClelland\cite{CME} and Marx and Szalay\cite{MS} with
regards to neutrinos.

{\bf Cold relics}, on the other hand, undergo freeze-out at temperatures
much less than their mass: $T_\chi \ll m_\chi$.  In this case their 
density is suppressed relative to the photon density by a Boltzmann factor,
so that at freeze-out $n_\chi \ll n_\gamma$.  This density, however, can be
deduced from the freeze-out condition itself
\begin{equation}
\Gamma_\chi=n_\chi\langle\sigma v\rangle_\chi
\biggr{|}_{\hbox{freeze-out}} = H
\simeq 1.7(g^*)^{1/2} T_\chi^2/M_{\rm P}~,
\end{equation}
where $g^*$ is the effective number of degrees of freedom at freeze-out
and $\langle\sigma v\rangle_\chi$ is the, thermally-averaged, annihilation
rate for the cold dark matter relic $\chi$.  In this case, the 
contribution to $\Omega_0$ of the relic depends both on this annihilation rate
and on the ratio $m_\chi/T_\chi$.  One finds, approximately,\cite{KT}
\begin{equation}
\Omega_\chi[{\rm CDM}] \simeq \frac{m_\chi}{T_\chi}
\left[\frac{10^{-27}{\rm cm}^3/{\rm sec}}
{\langle\sigma v\rangle_\chi\big|_{\hbox{freeze-out}}}\right]~,
\end{equation}
a formula first deduced by Zeldovich\cite{Zel}.  For a typical ratio
$m_\chi/T_\chi\sim 20$ one needs cross sections of $O(\sigma\sim 10^{-36}~
{\rm cm}^2)$.  These cross sections are of the typical strength of weak
interaction processes.  It is clearly intriguing that such cross sections could
have cosmological significance!

There are no simple formulas to describe the contributions to $\Omega$
of non-thermal relics, since these contributions depend in detail on the
dynamics.  In general, for non-thermal relics, the interactions are so
feeble that one has always $\Gamma_\chi \ll H$; that is, the relics are
never in thermal equilibrium.  For example, as we shall see later on,
axions with $m_a\simeq 10^{-5}~{\rm eV}$ can close the Universe
$(\Omega_{\rm axions} \simeq 1)$.  This means that the number density of
axions in this case is about $10^7$ times what it would be if axions were
thermal relics (i.e. had a $3^\circ~{\rm K}$ temperature).  So, if axions
are the dark matter in the Universe, they 
obviously had a highly non-thermal origin.

\section{Prospects of Neutrinos as Dark Matter}

Neutrinos are interesting candidates for dark matter since their
properties fit the required profile.  Furthermore, neutrinos are the only
dark matter candidates whose existence is confirmed experimentally!
Originally, neutrinos were thought to provide 
possible examples for both hot dark
matter and cold dark matter.  
Because of LEP, we know now that they can only
be HDM candidates.  Let me elaborate on this point.

Experimentally, one has evidence that the three known neutrinos,
$\nu_e,\nu_\mu$, and $\nu_\tau$, are quite light, with direct bounds on
their masses given by\cite{PDG}
\begin{equation}
m_{\nu_e} \leq 15~{\rm eV}~; ~~~~
m_{\nu_\mu} \leq 170~{\rm keV}~; ~~~~
m_{\nu_\tau} \leq 24~{\rm MeV}~.
\end{equation}
Because the freeze-out temperature for neutrinos is of order of 
$T_f\sim 1~{\rm MeV}$, so
as not to overclose the Universe $\nu_\mu$ and $\nu_\tau$ must have masses much
below these bounds.  Hence, it is perfectly conceivable that the known
neutrinos are the hot dark matter, contributing to $\Omega_0$ an amount
\begin{equation}
\Omega_\nu = \left[\frac{\sum_i m_{\nu_i}}{92~{\rm eV}}\right]
\frac{1}{h^2}~.
\end{equation}

If heavy neutrinos existed, with masses $m_{\nu_H} \gg T_f \sim 1~{\rm MeV}$,
they could be cold dark matter candidates because they have weak interactions.
It is straightforward, knowing the interaction rate of
neutrinos, to calculate their contribution to $\Omega$ as a function of the
neutrino mass.  The result is displayed in Fig. 8, taken from\cite{KT}.
This figure shows that $\Omega_\nu$ grows with $m_\nu$ up to around
the freeze-out temperature $T_f$
and then decreases rather rapidly for neutrino masses beyond this temperature.
However, the existence of further neutrino species with
$m_\nu \leq \frac{1}{2} M_Z$ is now excluded by measurements of the
$Z$-width at LEP, which, as we saw earlier, gives $N_\nu = 3$ to high
accuracy.  Thus the window for heavy neutrino CDM is closed.

\begin{figure}
\centerline{
\epsfig{file=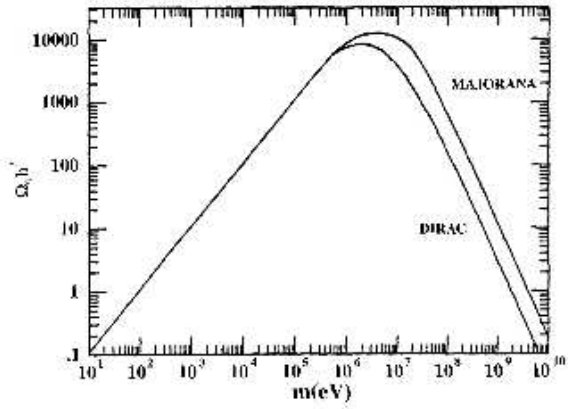,width=4in}
}
\caption[]{Contribution of neutrinos of different masses to $\Omega h^2$, from \cite{KT}}
\end{figure}

Although neutrinos are not cold dark matter candidates, they remain excellent
prospects for hot dark matter.  Nevertheless, because one knows that hot
dark matter alone cannot describe the power spectrum of density 
fluctuations\cite{Primack1},\footnote{HDM neutrinos, because they are so light,
have a free streaming length of $O(10~{\rm Mpc})$. As a result, they cannot account for
the formation of structure at small scales.}
one expects $\Omega_\nu < 1$.
Acceptable fits to the power spectrum of density fluctuations suggest
typically\cite{Primack1} $\Omega_\nu \simeq 0.2$.  Using $h^2 \simeq 0.3$,
if neutrinos are the HDM, this gives
\begin{equation}
\sum m_{\nu_i} \simeq 5-6~{\rm eV}~.
\end{equation}
This ratio is consistent with the bounds\cite{PDG} given in Eq. (46).
Unfortunately, however, there is as yet no direct particle physics evidence
that the known neutrinos have masses that satisfy Eq. (48).  Nevertheless,
there is tantalizing indirect evidence for neutrino masses (through hints of
neutrino oscillations) and this evidence is compatible with Eq. (48).
Because of its importance to the issue at hand, I review next some of this
information and its implications.

First, let me make a comment on prospects for improving the direct neutrino
mass bounds quoted in\cite{PDG}.  Clearly, even though the kinematical 
techniques that give the bounds on $m_{\nu_\mu}$ and $m_{\nu_\tau}$ can 
perhaps be
improved somewhat, 
there is no hope to directly measure masses in the few eV range
for these particles.  This is not so for $\nu_e$.  In fact, the tritium
$\beta$-decay experiments that lead to the bound of $m_{\nu_e} \leq
 15~{\rm eV}$
quoted by the Particle Data Group\cite{PDG}, actually all have
sensitivities of
order 1-2 eV!  The reason for the much weaker bound quoted, is that all the
latest precision experiments\cite{Npre} are plagued by an anomalous
unexplained {\bf excess} of events beyond the tritium beta decay endpoint.
This excess actually leads to an average mass-squared that is
{\bf negative} ($\langle m_{\nu_e}^2\rangle = 
(-27 \pm 20)~{\rm eV}^2)$.\cite{PDG}
Until this excess is understood, one cannot set a real bound for
$m_{\nu_e}$, although the potential sensitivity to eV masses is there.

In this context, one should mention a different piece of evidence that
suggests that $\nu_e$ itself cannot be the dominant form of the HDM.  This latter
constraint comes from double-beta decay, where searches for the neutrinoless
mode in $^{76}{\rm Ge}$ decay\cite{zbeta} lead to a limit on the effective
neutrino Majorana mass responsible for this process of \footnote{The
uncertainty in the bound of Eq. (49) reflects an uncertainty in the
calculation of the relevant nuclear matrix element.}
\begin{equation}
\langle m_\nu\rangle_{ee} \leq (0.5-1.5)~{\rm eV}~.
\end{equation}
Here $\langle m_\nu\rangle_{ee}$ is the sum 
of the neutrino masses entering in the process,
convoluted with the appropriate mixing matrix element coupling these neutrinos 
(and antineutrinos) to electrons:
\begin{equation}
\langle m_\nu\rangle_{ee} = \sum_i U^2_{ei} m_i~.
\end{equation}
If mixing of electrons to neutrinos other than $\nu_e$ is not large
$(U_{e1} \simeq 1)$, then Eq. (49) is also a bound on $m_{\nu_e}$.\footnote{
I am being a little sloppy here not distinguishing between weak interaction
eigenstates and mass eigenstates. Also, I am explicitly assuming that neutrinos
are Majorana particles.}  However, this is not a true bound because if
neutrinos are Dirac particles, the particle and antiparticle contributions
in Eq. (50) automatically cancel and $\langle m_\nu\rangle_{ee} \equiv 0$.

Fortunately, one can probe neutrino masses indirectly by looking for 
evidence for oscillations of one neutrino species into another.  If
neutrinos have mass, neutrinos can mix with one another and this mixing
can be revealed through neutrino oscillation experiments.  At present there
are a number of tantalizing hints arising from experiments looking for
neutrino oscillations which have an important bearing on the question
of neutrino mass.  To discuss these experiments, it is necessary first to
briefly discuss a bit of phenomenology.

If neutrinos have mass, the weak interaction eigenstates (the neutrinos
produced by weak interaction processes--e.g. $W^+\to e^+\nu_e$), are not the
same as the mass eigenstates (i.e. the observed particles of well defined
mass, denoted here by $\nu_i$).  However, these states are related by a 
unitary transformation, so that each $\nu_\ell~\{\ell = e,\mu,\tau\}$ can be
written as a superposition of the $\nu_i$:
\begin{equation}
\nu_\ell = \sum_i U_{\ell i}\nu_i~.
\end{equation}
Conventionally, one only examines the $2\times 2$ case, assuming that, as in the
quark case, the $3\times 3$ matrix will be nearly diagonal with 
dominant mixing among pairs of neutrinos.  In this case, for instance, Eq.
(51) for $\nu_e$ and $\nu_\mu$ just involves a simple orthogonal $2\times 2$
matrix:
\begin{equation}
\left(
\begin{array}{c}
\nu_e \\ \nu_\mu 
\end{array} 
\right) =
\left(
\begin{array}{cc}
\cos\theta & \sin\theta \\
-\sin\theta & \cos\theta
\end{array}
\right)
\left(
\begin{array}{c}
\nu_1 \\ \nu_2
\end{array}
\right)~.
\end{equation}
The mass eigenstates $\nu_i$ have the usual quantum mechanical evolution with
time:
\begin{equation}
|\nu_i(t)\rangle = e^{-iE_it}|\nu_i(0)\rangle~.
\end{equation}
Imagine then producing at $t=0$ a $\nu_e$ from a weak decay
\begin{equation}
|\nu(0)\rangle\equiv |\nu_e\rangle =
\cos\theta|\nu_1(0)\rangle + \sin\theta|\nu_2(0)\rangle~.
\end{equation}
At a later time, because the states $\nu_1$ and $\nu_2$ have different
masses, this state will evolve into a superposition of both
$|\nu_e\rangle$ and $|\nu_\mu\rangle = -\sin\theta|\nu_1(0)\rangle +
\cos\theta|\nu_2(0)\rangle$.  That is
\begin{eqnarray}
|\nu(t)\rangle &=& \cos\theta |\nu_1(t)\rangle + \sin\theta|\nu_2(t)\rangle 
\nonumber \\
&=& \cos\theta e^{-iE_1t}|\nu_1(0)\rangle + \sin\theta 
e^{-iE_2t}|\nu_2(0)\rangle~.
\end{eqnarray}

Using the above, it is easy to calculate the transition probability that an
initial $\nu_e$ state has oscillated into a $\nu_\mu$ state after a time
$t$:
\begin{eqnarray}
P(\nu_e \to \nu_\mu;t) &=& |\langle\nu(t)|\nu_\mu\rangle|^2 \nonumber \\
&=& \frac{1}{2} \sin^22\theta[1-\cos(E_1-E_2)t]~.
\end{eqnarray}
In all cases of interest $|p|\gg m_i$.  Hence
$E_i\simeq |p| + \frac{m_t^2}{2|p|}$, with $|p|\equiv E_\nu$ being
essentially the neutrino energy.  Also, in this case, the time $t$ in
Eq. (56) can just be replaced by the distance travelled (in units of the
speed of light): $t = L/[c]$.  Whence, one finds the following formula for
the probability that, as a result of neutrino mixing, 
an initial $\nu_e$ of energy $E_\nu$ has oscillated
after a distance $L$ into a $\nu_\mu$: 
\begin{eqnarray}
P(\nu_e\to \nu_\mu;L) &=& \sin^22\theta \sin^2
\left[\frac{(m_1^2-m_2^2)L}{4E_\nu}\right] \nonumber \\
&=&\sin^22\theta \sin^2\left[1.27 \frac{\Delta m^2({\rm eV}) L(m)}
{E_\nu({\rm MeV})}\right]~.
\end{eqnarray}
Of course,
\begin{equation}
P(\nu_e\to \nu_e;L) = 1-P(\nu_e\to \nu_\mu;L)~.
\end{equation}

One sees from the above formulas that the probability of oscillation is
sensitive to the mixing angle $\theta$.  Further, if one wants to probe a
particular $\Delta m^2$ range, then for a given neutrino energy $E_\nu$ there
are appropriate distances $L$ where the effect is maximum.  If the
neutrinos are not nearly degenerate, then the $\Delta m^2$ range one wants to
probe to find out whether neutrinos contribute significantly to the dark
matter problem ({\bf cosmologically} significant neutrinos) is $\Delta m^2 \sim
25~{\rm eV}^2$.  This is the goal of the CHORUS and NOMAD experiments at
CERN, which for
$\Delta m^2$ in this range hope to be sensitive to $\nu_\mu\to\nu_\tau$
oscillations as low as $\sin^22\theta_{\mu\tau} \geq 10^{-3}$.

Up to now the CERN experiments have only given limits.\cite{CEPN}  However, 
in other regions of $\Delta m^2$ there are various hints of 
neutrino oscillations.  In fact, there is an embarrassment of riches!
The LSND experiment\cite{LEND} sees a signal of $\bar\nu_\mu\to\bar\nu_e$
oscillations in a narrow range in $\Delta m^2$ around (0.2-2) ${\rm eV}^2$, with
$\sin^22\theta \simeq 10^{-2}~{\rm eV}^2$.  The ratio of $\nu_\mu$ to $\nu_e$
neutrinos, observed in large underground experiments, 
arising from decay processes in the atmosphere shows a 
deficit\cite{atmospheric} from 
what is expected.
This atmospheric anomaly can be interpreted either as
being due to $\nu_\mu\to\nu_\tau$ or $\nu_\mu\to\nu_e$ oscillations, with
$\Delta m^2\sim (0.3-3)\times 10^{-2}~{\rm eV}^2$ and large mixing angles
$\sin^22\theta\simeq O(1)$.\footnote{Very recent data from 
SuperKamiokande\cite{SueprK} favors a lower range for $\Delta m^2\sim
(0.1-1)\times 10^{-2}$, while the negative results from the Chooz\cite{choose}
reactor experiment now excludes the $\nu_\mu\to\nu_e$ oscillation option
for explaining the atmospheric anomaly.}  Finally, experiments measuring
the flux of solar neutrinos, also show a dearth of neutrinos compared to the
predictions of the, so called, standard solar model.\cite{SSM}  One
can reconcile the observations of all of these 
solar neutrino experiments by appealing to
$\nu_e\to\nu_\mu$ or $\nu_e\to\nu_\tau$ neutrino oscillations, which are
enhanced in matter by the so-called MSW mechanism,\cite{MSW} provided that
$\Delta m^2\sim (0.3-1.2)\times 10^{-5}~{\rm eV}^2$ with rather small
mixing: $\sin^22\theta\sim (4-10)\times 10^{-3}$.\cite{uter}

\begin{figure}
\centerline{
\epsfig{file=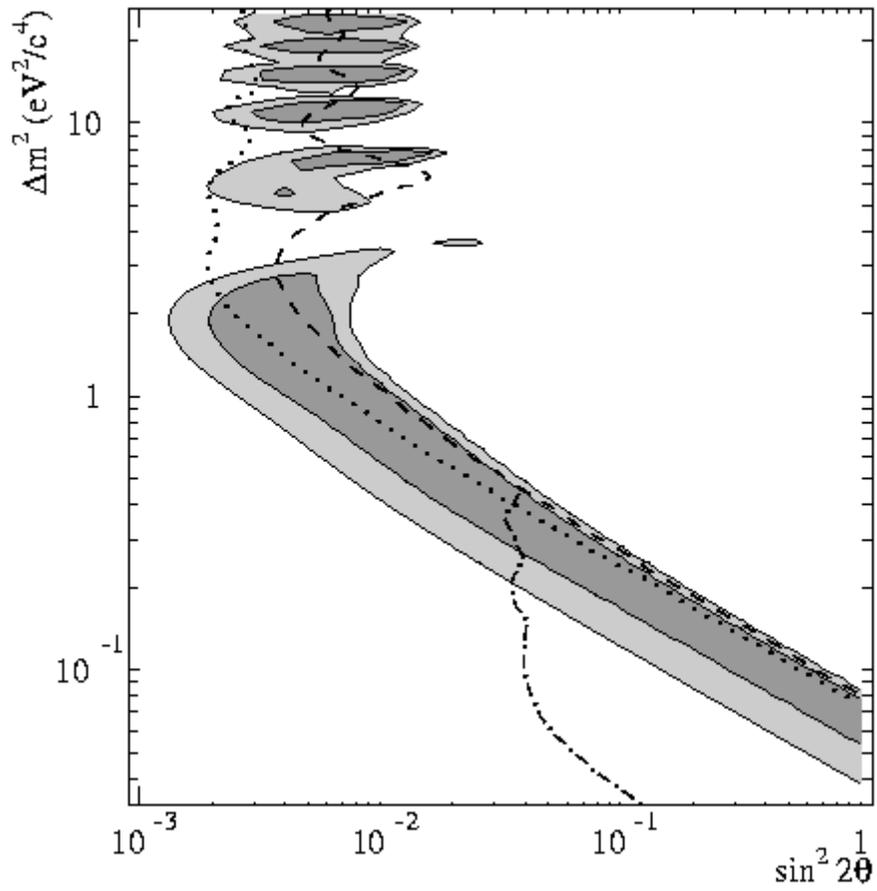,width=5in}
}
\caption[]{Signal region for the LSND experiment, along with exclusion limits from other experiments \cite{LEND}}
\end{figure}

Because we know of only three neutrino species, these hints cannot all be
true, since we have at most only two mass differences!\footnote{It is 
possible not to discard any experimental 
hints if one assumes that, in
addition to $\nu_e,\nu_\mu,$ and $\nu_\tau$, there is an extra {\bf sterile}
neutrino $\nu_s$.  
Then one of the experimental results--the solar anomaly--can
be interpreted as a $\nu_e\to\nu_s$ oscillation\cite{Caldwell}.}
Even if one were to eliminate one of the hints (LSND perhaps--since,
as Fig. 9 make clear, the allowed region is almost ruled out by other
negative findings), because the favored mass differences are small, it
appears that to have cosmologically significant neutrinos one must have
near mass degeneracy.  For example, 
the pattern $m_{\nu_1}\simeq m_{\nu_2}\simeq m_{\nu_3}\simeq 1-2~
{\rm eV}$, with $\Delta m^2_{12}\sim 10^{-5}~{\rm eV}^2$; $\Delta m^2_{23}\sim
10^{-3}~{\rm eV}^2$ would explain the solar and atmospheric anomaly.  If this were
really the case, then cosmologically significant neutrinos would produce
{\bf no signal} in CHORUS and NOMAD!\footnote{In this scenario,
one has to worry about the double-beta decay limits, since these provide
effective electron neutrino masses $\langle m_\nu\rangle_{ee}$ precisely
in this range.}  Hopefully, in the next five years with upcoming neutrino
oscillation experiments (as well, perhaps, with some clarification in the
direct tritium $\beta$-decay experiments) one should be able to sort out this
somewhat confusing situation, thereby arriving at a better understanding of 
whether or not neutrinos can 
contribute to the dark matter in the Universe.\footnote{The study of the
angular power spectrum of the cosmic background radiation can also provide
information on this issue, as massive neutrinos can affect this spectrum
differently depending on their mass.\cite{Burns}}

Before closing this section, I would like to make an important theoretical
point.  Neutrino masses in the eV and sub-eV ranges are very interesting from
a particle physics point of view, since they are most likely a signal for
a new large mass scale.  In general, because neutrinos are neutral, they can 
have both {\bf Dirac} - particle/anti-particle---and {\bf Majorana} - 
particle/particle masses.  That is, one can write
\begin{equation}
{\cal{L}}_{\rm mass} = -m_D\bar\nu_{\rm L}\nu_{\rm R} - \frac{1}{2}
m^{\rm R}_M\nu^T_{\rm R}C\nu_{\rm R} - \frac{1}{2} m^{\rm L}_M
\nu^T_{\rm L}C\nu_{\rm L} + h.c.~,
\end{equation}
where $C$ is a charge conjugation matrix.\cite{RDPB}  Because $\nu_{\rm L}$
is part of an SU(2) doublet, while $\nu_{\rm R}$ (if it exists!) is part of
an SU(2) singlet, it is clear that in the standard model $m_M^{\rm R},m_D$
and $m_M^{\rm L}$, respectively, carry effective SU(2) quantum numbers of
0, 1/2 and 1.  In particular, while $m_M^{\rm R}$ can be a totally independent
mass parameter, $m_D$ and $m_M^{\rm L}$ must be proportional to the
vacuum expectation value of an SU(2) doublet and triplet field, respectively.
We know, as a result of the experimentally very
successful interrelation between $M_W$ and $M_Z$:  $M_Z^2\cos^2\theta_W=
M^2_W$, 
that what causes the breakdown of
the electroweak theory through
its VEV transforms dominantly as
an SU(2) doublet. Thus, if $\nu_{\rm R}$ exists,
we expect $m_M^{\rm L}\ll m_D$, with  $m_D\sim m_\ell$---the mass of 
the corresponding lepton.

It was realized long ago by Yanagida\cite{Yanagida} and
Gell-Mann, Ramond and Slansky,\cite{GRS} that if the Majorana mass of the right-handed neutrinos $m_M^{\rm R}$ is very
large, $m_M^{\rm R} \gg m_D$,
the above scenario produces very tiny neutrino masses.
If one neglects altogether $m_M^L$,
$m_M^L\simeq 0$, one has a $2\times 2$ neutrino mass matrix of the form
\begin{equation}
{\cal{M}} = 
\left(
\begin{array}{cc}
0 & m_D \\  
m_D & m_M^{\rm R}
\end{array}
\right)~.
\end{equation}
This matrix has a very heavy neutrino, mostly $\nu_{\rm R}$, with mass
$m_M^{\rm R}$ and a very light neutrino, mostly $\nu_{\rm L}$, with mass
\begin{equation}
m_\nu \simeq \frac{m_D^2}{m_M^{\rm R}} \sim
\frac{m_\ell^2}{m_M^{\rm R}}~.
\end{equation}
This, so-called, see-saw mechanism \cite{Yanagida} can produce eV neutrino masses provided
$m_M^{\rm R}$ is sufficiently large (e.g. for $m_{\nu_\tau}$ one has eV
neutrinos associated with the tau if $m_M^{\rm R}\sim 10^{10}~{\rm GeV}!$).
Thus detecting light neutrino masses is tantamount to discovering a
large scale---the scale responsible for the right-handed neutrino Majorana 
mass.\footnote{I should comment that even if there were no $\nu_{\rm R}$---
something I consider unlikely---the presence of eV neutrino masses again,
most likely, reflects another large mass scale.
For instance, without $\nu_{\rm R}$ one
can get a Majorana mass for $\nu_{\rm L}$ by using a doublet Higgs field
twice to make a triplet.  Such interactions are non-renormalizable, but could
arise effectively from some GUT interactions\cite{Weinberg} and are scaled by $1/M_{\rm GUT}$.  A formula like $m_\nu = m_M^{\rm L} \sim 
\langle\Phi\rangle^2/M_{\rm GUT}$, with $\langle\Phi\rangle \sim 300~{\rm GeV}$
gives eV neutrino masses for $M_{\rm GUT}\sim 10^{14}~{\rm GeV}$.}

\section{Supersymmetric Candidates for Dark Matter}

Supersymmetric extensions of the Standard Model for the strong and electroweak
interactions provide excellent candidates for cold dark matter. 
Supersymmetry, as is well known\cite{susy}, is a fermion-boson
symmetry.  Thus, if it were a true symmetry of nature, we 
would expect a doubling
of all the degrees of freedom.\footnote{For technical reasons, connected to
the cancellation of anomalies, one needs also to double the number of Higgs
doublets, as well as provide appropriate fermionic partners to these
states.}  Thus, in these extensions of the Standard Model,  
there is a plethora of undiscovered particles.  
Some of these particles turn out to be good 
dark matter candidates.  This is plausible because
a supersymmetric extension of the standard model preserves the strength of
the couplings.  For
instance, the supersymmetric vertex joining a squark (the scalar
partner of the quark), a quark and a gaugino (the spin-1/2 partner of a
gauge boson) has the same strength as the quark-quark-gauge boson vertex.
As a result,
(some) of the interaction cross sections for supersymmetric (SUSY)
particles will have the strength of the weak interactions
(provided these particles are not much heavier than the weak bosons)
and thus will
satisfy Zeldovich's criteria for cold relics, Eq. (45).

Most supersymmetric extensions considered contain a discrete symmetry, called
$R$-parity:
\begin{equation}
R = (-1)^{3{\rm B+L}+2S}~,
\end{equation}
which is +1 for particles and -1 for sparticles.  If $R$ parity is conserved,
then the lightest supersymmetric particle, the LSP, by necessity is stable.
If it is neutral, as is usually assumed to be the case to avoid cosmological
difficulties associated with their luminosity,\cite{Ellis} the
LSP provides an excellent candidate for cold dark matter.

We know that if supersymmetry exists it must be broken in nature.  
Otherwise, the masses of the supersymmetric partners, $\tilde m$, would be the same as 
that of the ordinary 
particles, $m$, in gross contradiction with experiment.  However, we
do not know really how supersymmetry breaks down. As a result, which
particle is the LSP is model dependent.  Nevertheless, one can make some
general observations.  

There are three important scales associated with
SUSY breaking.  The first of these is, obviously, the masses of the
sparticles themselves, $\tilde m$.  The second is the scale,
$\Lambda$, which is associated with the spontaneous breaking of supersymmetry.
This is assumed to occur in a, so called, hidden sector, separated from the
ordinary interactions of particles.\footnote{Separating the
process of supersymmetry breaking from ordinary matter is necessary  to avoid contaminating
ordinary matter with interactions we have not yet seen.}  The last scale, $M$,
is the scale associated with whatever phenomena acts as the messenger
connecting the hidden sector with ordinary matter.  This connection is
shown pictorially in Fig. 10.  Both $\Lambda$ and $M$ are model dependent, but
one expects the masses of the sparticles to be of order
\begin{equation}
\tilde m \sim \frac{\Lambda^2}{M}~.
\end{equation}

\begin{figure}
\centerline{
\epsfig{file=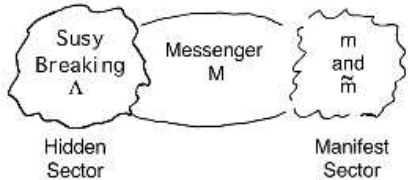,width=3in}
}
\caption[]{Scales associated with SUSY breaking}
\end{figure}

It is an attractive possibility that
supersymmetry resolves the naturalness problem of electroweak
symmetry breaking, related to why the scale of electroweak breaking
$v \sim 250~{\rm GeV}$ is so much less than the Planck mass. 
For this to be the case, SUSY states cannot themselves have
masses much bigger than $v$.  Hence, it is generally assumed that supersymmetric
partners must themselves be of mass $\tilde m \sim v$.\cite{neutralinos}
Thus $\Lambda$ and $M$, the parameters associated with supersymmetry breaking,
are constrained physically to produce $\Lambda^2/M\sim v$.  

With this 
constraint in mind, two main scenarios have emerged, with each scenario
producing a different LSP.  The first scenario arises out of supergravity
models,\cite{AN} where the hidden sector is connected to the ordinary sector
by gravitational interactions.  Here $M\sim M_{\rm P}$ and thus
$\Lambda\sim 10^{11}~{\rm GeV}$.  In the second scenario\cite{DN} the
messenger sector is 
associated with gauge interactions with a scale around
$M\sim 10^6~{\rm TeV}$.  To get $\Lambda^2/M \sim v$ necessitates then a much
lower scale $\Lambda$ of spontaneous supersymmetry breaking, $\Lambda\sim
10^3~{\rm TeV}$.

In both scenarios the gravitino, the spin-3/2 supersymmetric partner
of the graviton, becomes massive as a result of the spontaneous breaking
of supersymmetry, with a mass of order
\begin{equation}
m_{3/2} \sim \frac{\Lambda^2}{M_{\rm P}}~.
\end{equation}
From the above,
one sees that in the supergravity scenario, the gravitino mass is 
also of $O(v)$---typical of
all the other masses of the supersymmetric partners.  Thus, in this case, it is
generally
assumed that the gravitino is {\bf not} the LSP, but that 
the LSP is a {\bf neutralino}.  This is the lightest spin-1/2 partner
of the neutral bosonic particles in the theory---the two gauge bosons,
$\gamma$ and $Z$, and the two neutral Higgs bosons $H_{\rm u}$ and 
$H_{\rm d}$.  In general, the neutralino is some particular superposition of
all these spin-1/2 partners
\begin{equation}
\chi = a_\gamma\tilde \gamma + a_Z\tilde Z + a_{\rm u}\tilde H_{\rm u} +
a_{\rm d}\tilde H_{\rm d}~,
\end{equation}
where the $a_i$ are model dependent coefficients.  In contrast, in
the scenario where the messenger are gauge interactions, with
$M\sim 10^6~{\rm TeV}$ and $\Lambda \sim 10^3~{\rm TeV}$, the gravitino is
extraordinarily light,
\begin{equation}
m_{3/2} \sim\frac{\Lambda^2}{M_{\rm P}} \sim O({\rm KeV})
\ll \tilde m~,
\end{equation}
and is the LSP.  While neutralinos, with $m_\chi\sim O(10-10^3~{\rm GeV})$,
are typical cold dark matter relics, gravitinos, with $m_{3/2} \sim
O({\rm KeV})$, act as warm dark matter.  I discuss both of these cases, in
turn.

The typical supergravity model\cite{AN} which gives neutralino CDM is
characterized by a set of universal soft supersymmetry breaking parameters,
specified at the scale $\Lambda$.  In addition, it contains as a parameter 
the vacuum expectation ratio
between the $H_{\rm u}$ and $H_{\rm d}$ Higgs bosons:
$\tan\beta = \langle H_{\rm u}\rangle/\langle H_{\rm d}\rangle$.  The,
so called, minimal supersymmetric standard model (MSSM)\cite{MSSM}
 has actually only 2
additional parameters, besides $\tan\beta$, which determine the LSP mass,
$m_\chi$, and the coefficients $a_i$ in Eq. (65).  These are the common
mass, $m_{1/2}$, of all the gauginos and a mass parameter $\mu$ characterizing
the supersymmetric coupling between the $H_{\rm u}$ and $H_{\rm d}$
supermultiplets.  However, even in this minimal model, the actual contribution
of the neutralino LSP to $\Omega_{\rm CDM}$ depends on the neutralino
annihilation cross sections.  These, in turn, depend on other model
parameters, the universal soft breaking mass $m_o$ given to all the scalars
and certain coefficients ($A$ and $B$) which typify the strength of
trilinear and bilinear soft interaction terms.\cite{MSSM}
\begin{figure}
\centerline{
\epsfig{file=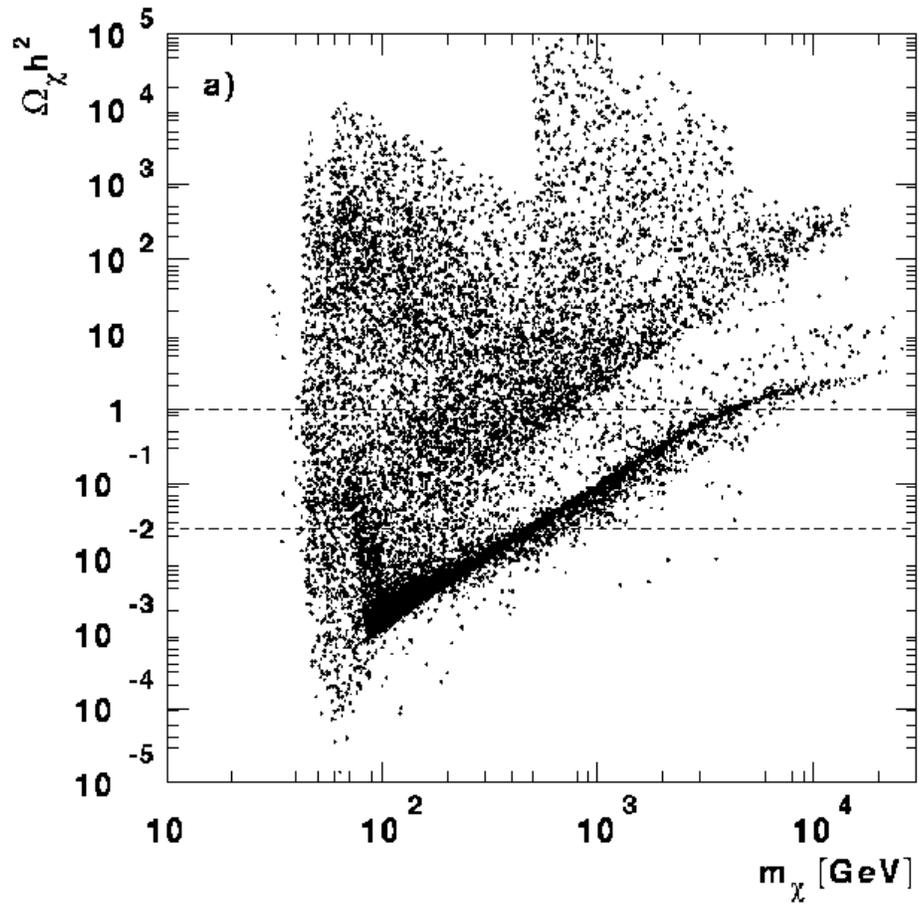,width=5in}
}
\caption[]{Contribution to $\Omega_{\chi}$ of various neutralino LSP models, from \cite {Gondolo}}
\end{figure}

As a result, even in the MSSM,
there is a large region of parameter space which produces a neutralino
LSP which potentially could be 
the cold dark matter in the universe.  Typically, what one requires for a
viable model is that $\Omega_\chi h^2 = 0.2 \pm 0.1$.  As can be seen in Fig. 
11, there are plenty of models (each represented by a dot) which have
$50~{\rm GeV} \leq m_\chi \leq 200~{\rm GeV}$ and lead to $\Omega_\chi h^2$ in the
desired range, provided that $\tan\beta$ is small and the resulting
pseudoscalar Higgs mass $m_A$ is large ($m_A \sim 500~{\rm GeV}$).\cite{Bottino}
\cite{Gondolo}
In general, an LSP much below about 50 GeV runs into trouble with the
negative results from LEP on Higgs searches, as well as on the direct production
of supersymmetric pairs.\cite{Jellis}  Thus there are regions in
parameter space that are already excluded, serving 
to rule out some potential CDM models.
Clearly the discovery of an LSP would have an enormous impact on the CDM
question, much reducing the parameter freedom one still has now, even for the
simplest models.

In gauge mediated supersymmetry breaking models, in contrast, the gravitino is
the LSP.  Here one has much less freedom since there are not that many
parameters to vary.  Gravitino interactions scale as $1/\Lambda^2$, and so are
typically very much weaker than weak interactions
\begin{equation}
\sigma_{3/2} \sim \left[\frac{1~{\rm TeV}}{\Lambda}\right]^4
\sigma_{\rm weak} \ll \sigma_{\rm weak}~.
\end{equation}
As a result, the freeze-out of these interactions occurs at an earlier
epoch in the Universe, when there were more thermal degrees of freedom. Thus, gravitinos have a smaller abundance compared to neutrinos
of the same mass.\cite{PP}  Typically, one finds
\begin{equation}
\Omega_{3/2}h^2 \simeq \left[\frac{m_{3/2}}{1~{\rm KeV}}\right]
\left[\frac{100}{g^*(T_f)}\right]~,
\end{equation}
where $g^*(T_f)$ is the number of degrees of freedom at freeze-out.
Because it takes gravitinos of mass of order 1 KeV
 to close the Universe,
cosmologically significant gravitinos have a smaller Jean's mass than 
neutrinos
\begin{equation}
M^{\rm Jeans}_{3/2} \sim \frac{M^3_{\rm P}}{m^2_{3/2}} \sim 10^{12}~M_\odot~.
\end{equation}
Thus, in contrast to neutrinos, the gravitino free-streaming 
length is rather small, of order
$\lambda_{3/2} \sim 1~{\rm Mpc}$, much closer
to that of cold dark matter.  Hence, gravitinos are typical warm dark 
matter---matter that is relativistic at decoupling but does not form only
large structures.  Indeed, as I just mentioned, the spectrum of density
fluctuations for gravitino dark matter is quite similar to that of cold
dark matter.  This is seen clearly in Fig. 12, from a recent study of Borgani,
Masiero, and Yamaguchi.\cite{BM}

\begin{figure}
\centerline{
\epsfig{file=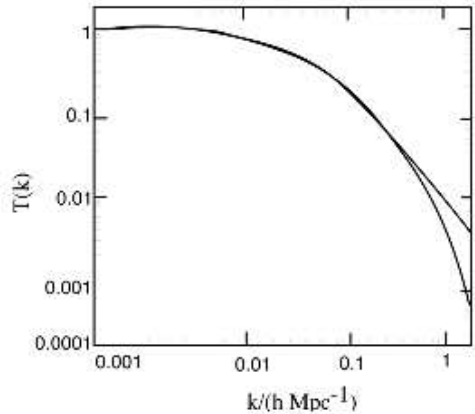,width=3.5in}
}
\caption[]{Transfer function for gravitinos and CDM showing their similarity, from \cite{BM}}
\end{figure}

Gravitinos, in my view, are not a particularly attractive form of dark matter,
as to get the needed $\Omega_{3/2}$ one needs to have the gravitino mass
($m_{3/2} \sim \Lambda^2/M_{\rm P}$) finely tuned around a KeV.  But this
is not the worse trouble!  Because of its extremely tiny interaction cross
section [cf Eq. (67)] gravitino dark matter does not have any hope to be detected ever.\footnote{Besides the tiny cross section, KeV gravitinos also
give too little energy of recoil.}  In contrast, if the CDM is due to a
neutralino LSP, in principle, it may be detectable by experimental
means.

Calculation of the rates expected in low background experiments
(for instance, those using a $^{73}{\rm Ge}$ detector of sufficient mass), 
depend both on the
density of LSPs in our galaxy and on the neutralino-nucleon scattering
cross section.  This latter cross section depends again on the various
parameters in the supersymmetric model.  Except for very light nuclei, it
turns out that scalar exchange dominates, since it leads to coherent
scattering of the neutralinos on the target nuclei, so that
$\sigma_{\chi A} \sim A^2$.  Fig. 13 shows that the expected rates 
of neutralino CDM for a
$^{73}{\rm Ge}$ detector are 
of the order of $10^{-2}-10^{-3}$ events/Kg-day.  Given that
present-day detectors (e.g. CDMS \cite{CDMS}) are operating with at best one
Kg of Ge, one is still looking for a factor of $10^2-10^3$ improvement
to have some hope of detecting a potential signal for neutralino cold
dark matter.
This is a daunting, but perhaps not impossible, task.  As I said earlier, the
experimental observation of a neutralino LSP in a particle physics experiment
would give enormous impetus to the lofty goal of direct dark
matter detection!

\begin{figure}
\centerline{
\epsfig{file=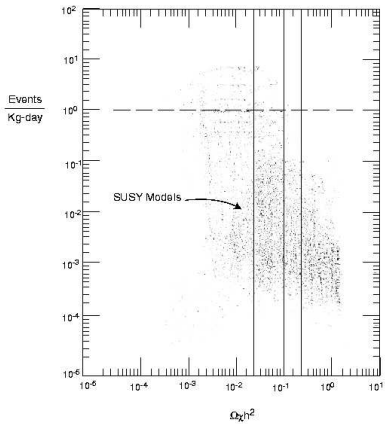,width=2.5in}
}
\caption[]{Expected rates from neutralino dark matter, from \cite{Bottino}}
\end{figure}

\section{Axions as CDM Candidates}

Axions are pseudo Goldstone bosons associated with a spontaneously broken
global chiral symmetry, $U(1)_{\rm PQ}$, introduced to ``solve" the,
so called, strong CP problem.\cite{PQ}  The Lagrangian of the electroweak
and strong interactions, in general, possesses an effective interaction
involving the gluon field strengths, $G_a^{\mu\nu}$ and their duals
$\tilde G_a^{\mu\nu}$:
\begin{equation}
{\cal{L}}_{\rm eff} = \bar\theta \frac{\alpha_s}{8\pi} G_a^{\mu\nu}
\tilde G_{a\mu\nu}~.
\end{equation}
This interaction breaks P, T, and CP and produces a very large neutron
electric dipole moment unless the parameter $\bar\theta$ is very 
small.\footnote{One finds $d_n\simeq 10^{-16}\bar\theta~{\rm ecm}$\cite{RDP}
and hence one needs to have
$\bar\theta\leq 10^{-10}$ to respect the strong experimental
bounds on $d_n$.\cite{PDG}}  This is the strong CP problem---why is
$\bar\theta$ so small?  The imposition of an additional global chiral
symmetry on the standard model suggested by Quinn and myself,\cite{PQ}
essentially serves to replace the $\bar\theta$ parameter by a dynamical
field---the axion field.\cite{WN}  Instead of the CP violating interaction (70)
one now has instead, a CP-conserving effective interaction of the axion 
field $a(x)$ with
the gluons:
\begin{equation}
{\cal{L}}_{\rm PQ} = \frac{a}{f} \frac{\alpha_s}{8\pi}
G_a^{\mu\nu} \tilde G_{a\mu\nu}~,
\end{equation}
where $f$ is a scale associated with the spontaneous breakdown of
$U(1)_{\rm PQ}$.

The axion is the Nambu-Goldstone boson associated with the spontaneous
breakdown of the $U(1)_{\rm PQ}$ symmetry.  However, because this symmetry
has a chiral anomaly---reflected in the appearance of the interaction 
(71)---the axion is not truly massless but acquires a small mass.\footnote{The
interaction (71) produces for the axion field an effective potential which
dynamically adjusts so as to cancel the $\bar\theta$ parameter.  This
potential also has a non-vanishing second derivative at its minimum,\cite{RDP}
corresponding to the axion mass.}  This mass is slightly model-dependent,
but is of order
\begin{equation}
m_a \sim \frac{m_\pi f_\pi}{f} \sim
\left(\frac{6\times 10^6}{f[{\rm GeV}]}\right)~{\rm eV}~.
\end{equation}
One sees that, for large $f$, axions are very light.  Since all couplings of
the axion scales as $1/f$, these particles, if they exist, are also very
weakly coupled.  Although axions are not stable since they can decay into
two photons, the lifetime for the process $a\to 2\gamma$ scales as
$\tau\sim f^5$\cite{RDP} and becomes enormous for large $f$.

Quinn and I\cite{PQ} made the natural assumption that the scale of
$U(1)_{\rm PQ}$ breaking coincided with the electroweak scale, $f\sim v$.
Unfortunately, these weak-scale axions have been ruled out 
experimentally.\cite{RDP}  If $f$ is not of $O(v)$, it turns out that
astrophysics constrains $f\gg v$.  This is easy to understand. Axions
provide an extremely efficient way to cool down stars, completely
affecting their evolution.  Only if $f\geq 5\times 10^9~{\rm GeV}$, are
axion couplings sufficiently weak so as not to run into trouble with a
variety of astrophysical observations---ranging
from the evolution of red giants, to
properties of the observed neutrino pulses from SN 1987a.\cite{Raffelt}
For $f\geq 5\times 10^9~{\rm GeV}$, the axions are so light, so weakly coupled,
and so long-lived to be effectively ``invisible".\cite{invisible}  However,
these invisible axions have potential cosmological consequence, and
they prove to be
interesting cold dark matter candidates!  Let me review the arguments
for this.\cite{acb}

Axions are typical non-thermal relics, since their properties change as
the Universe evolves.  At the $U(1)_{\rm PQ}$ phase transition, which occurs
when the Universe's temperature $T\sim f$, axions are produced as real
Nambu-Goldstone bosons $(m_a=0)$.  At such high temperatures the axion 
potential due to QCD is ineffective and the $\bar\theta$ interaction of
Eq. (70) is not cancelled out.  As the Universe cools towards temperatures
of order of the QCD-scale $\Lambda_{\rm QCD},~T\sim\Lambda_{\rm QCD}$, two
things happen: the axion potential turns on, serving to cancel $\bar\theta$,
and the axion acquires its mass, which is of $O(\Lambda^2_{\rm QCD}/f)$.
This relaxation of the axion field to its present configuration, however,
happens in an oscillatory way.  The energy density associated with these
oscillations, as we shall see, acts as cold dark matter.\cite{acb}

The $\bar\theta$ parameter in Eq. (70) can be thought of as an effective VEV
for the axion field:  $\langle a\rangle = \bar\theta f$, with the correct
vacuum state driving $\langle a\rangle\to 0$.  In the early Universe at
$T\sim f$, in this language, the axion field has an effective vacuum
expectation $\langle a\rangle = \bar\theta f$.  As the temperature 
lowers towards $T\sim\Lambda_{\rm QCD}$, the QCD potential for the axion
turns on and $\langle a\rangle$ is driven to zero.  One can study the time
evolution of $\langle a\rangle$ by studying the
equation of motion for the axion field in the expanding Universe:\cite{acb}
\begin{equation}
\frac{d^2\langle a\rangle}{dt^2} + 3\frac{\dot R(t)}{R(t)}
\frac{d\langle a\rangle}{dt} + m_a^2(t)\langle a\rangle = 0~.
\end{equation}
It is clear from the above that the effect of the expansion 
of the Universe is to provide a
drag term for $\langle a\rangle$.  At early times, or high temperatures,
the axion mass vanishes and $\langle a\rangle$ is fixed to its initial
value $\langle a\rangle = f\bar\theta$.  
When the axion mass $m_a(t)$ begins to
turn on, as the Universe's temperature cools towards $T\sim\Lambda_{\rm QCD}$,
$\langle a\rangle$ undergoes damped oscillations about
$\langle a\rangle=0$.

I shall not try to sketch here the computation of the effective energy density
associated with these oscillations of $\langle a\rangle$, but refer to
Ref.\cite{RDP} for an elementary discussion.  I quote, however, the
result of a recent detailed calculation\cite{Turner} which gives the
contribution to $\Omega_0$ of these oscillations. One finds:
\begin{equation}
\Omega_ah^2 = C\left[\frac{f}{10^{12}~{\rm GeV}}\right]^n
\bar\theta^2~.
\end{equation}
Here $C$ is a constant of $O(1)$ which depends on the details of the QCD
phase transition, while the exponent $n$ is near unity, 
$n=1.18$.  One sees that if the initial value
for $\langle a\rangle/f = \bar\theta$ is of $O(1)$---as one may expect
naively---then these oscillations of the axion VEV can close the Universe
if $f\sim 10^{12}~{\rm GeV}$.  Because what is oscillating is
$\langle a\rangle$, these oscillations correspond physically to coherent,
zero momentum, oscillations of the axion field.  Since $\vec p_a=0$,
axion oscillations are prototypical cold dark matter.

From the above, it appears that coherent axion oscillations can give rise
to $\Omega = 1$ provided $f\simeq 10^{12}~{\rm GeV}$ or $m_a\simeq
6\times 10^{-6}~{\rm eV}$.  This is predicated on having an initial
misallignment angle $\bar\theta\sim O(1)$.  However, Linde\cite{Linde} has
argued that in inflationary cosmology, with the reheating temperature
$T_{\rm reheating} <f$ so that there is not a post-inflationary
$U(1)_{\rm PQ}$ phase transition, there is no reason why the misallignment
angle cannot be very small:  $\bar\theta^2\ll 1$.  In this case one could
have $\Omega_a\sim O(1)$ for smaller axion masses (or
$f \gg 10^{12}~{\rm GeV}$):
\begin{equation}
\Omega_a\sim O(1)~~\hbox{if}~~m_a\simeq 6\times 10^{-6}\bar\theta^2~{\rm eV}~.
\end{equation}

There are other arguments, however, which suggest that axion masses much
heavier than $m_a\simeq 6\times 10^{-6}~{\rm eV}$ can close the Universe.
These arguments apply in inflationary scenarios where the reheating temperature
$T_{\rm reheating} >f$.  In this case, one must worry about axionic strings
formed at the $U(1)_{\rm PQ}$ phase transition.  The decay of these strings
into axions also contributes to the Universe's energy density and this
contribution can dominate that due to coherent axion oscillations.\cite{Davis}
Unfortunately, there is considerable controversy on this point, with some
authors---notably P. Sikivie and collaborators\cite{HS}---obtaining
$\Omega_{\rm string~decay} \sim \Omega_{\rm oscillation}$, with 
others\cite{BS} deducing $\Omega_{\rm string~decay}\gg
\Omega_{\rm oscillation}$.  If one were to believe this latter estimate, then
one obtains $\Omega_a\sim O(1)$ for axion masses as heavy as
$m_a \sim 10^{-4}~{\rm eV}$.  These masses are perilously close to the mass
range excluded by astrophysics, \cite{Raffelt} $m_a \geq 10^{-3}~{\rm eV}$,
corresponding to $f<5\times 10^9~{\rm GeV}$.

\begin{figure}
\centerline{
\epsfig{file=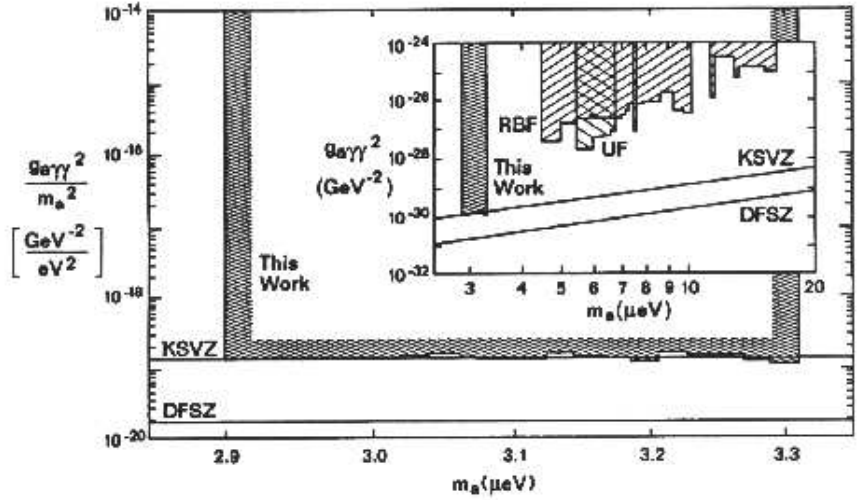,width=5in}
}
\caption[]{Results of invisible axion searches }
\end{figure}

These controversies may be resolved experimentally if axions are the dark
matter in the Universe (and hence are also the dominant form of the dark
matter in our galaxy!).  The basic idea for these experiments is due to 
Sikivie\cite{Sikivie} and uses the fact that axions couple to the electromagnetic field in a way analogous to how they couple to gluons
[cf. Eq. (71)]:
\begin{equation}
{\cal{L}}_{a\gamma\gamma} = g_{a\gamma\gamma}a\vec E\cdot\vec B
\end{equation}
with $g_{a\gamma\gamma}\sim 1/f$.  Because of Eq. (76) axions in our
galactic halo in the presence of a strong magnetic field can be
resonantly converted into photons in an appropriate cavity.  Experiments are
presently underway at both the Lawrence Livermore Laboratory\cite{LLL} and at Kyoto University\cite{Kyoto}
which are sensitive to ``standard" invisible axions if they are the
dominant form of dark matter.\footnote{``Standard" in this context means
invisible axions with an initial misallignment angle $\bar\theta\sim O(1)$
and ones where coherent axion oscillations dominate the energy density 
contribution.}  Fig. 14 shows recent results from the Livermore experiment \cite{LLL}
in the $g^2_{a\gamma\gamma}-m_a$ plane (along with
some regions already excluded by some initial pioneering 
experiments)\cite{pioneer} 
and the theoretical expectations of invisible axion models. The hope is that when both the Livermore and Kyoto experiments are completed, in 3-5 years, one will know whether axions are, or are not, an important component of the dark matter in the Universe 

\section{Perspectives on Dark Matter}

It is useful at this stage to try to bring some perspective on the issue of
dark matter from a particle physics point of view.  As we saw, particle
physics provides an interesting array of dark matter candidates.  Among these,
it appears that
perhaps the neutralino LSP is the particle physics relic which is the most
plausible dark matter candidate.  In the simplest supersymmetric extension
of the standard model, the MSSM,  
there is a rather large range in parameter space which gives rise to a neutralino LSP that has 
$\Omega_\chi h^2\sim O(1)$.  In contrast, both for axions, gravitinos and
neutrinos, 
the critical density in the Universe obtains only for some specific values of
the parameters characterising these excitations (e.g. for axions one needs the
scale of $U(1)_{\rm PQ}$ breaking, $f$, to be of $O(10^{12}~{\rm GeV})$).

Although, on the 
face of it, the above argument seems very reasonable, I am not sure it is 
totally compelling.  For instance, in a similar vein one could
argue also that having $\Omega_B = 0.05$ is unnatural, since it requires a
peculiar tuning of the nucleon mass!  I believe a more sensible point of view
to take is the following.  Of all the cosmological scenarios,
the inflationary scenario for the Universe appears to make the most sense.
If this scenario is correct, then $\Omega = 1$ is a boundary condition one
should seriously impose as a constraint on the {\bf sum} of all the
particle species which are important in the Universe today.  That is, we
should demand that\footnote{In principle, one of the $\Omega_i$ could be the
contribution from a cosmological constant.}
\begin{equation}
1 = \Omega = \sum_1 \Omega_i~.
\end{equation}
The particular weight of each of the components $\Omega_i$ in Eq. (77)
is a reflection of intrinsic particle physics properties.  The only
cosmological constraint is that the sum of the $\Omega_i$ must add up to
unity.  So, if particle physics arguments lead to $f\simeq 10^{12}~{\rm GeV}$,
or $m_{\nu_\tau} \simeq 5~{\rm eV}$, then that particular component will be
important in the sum appearing in Eq. (77).  From this point of view,
``what you see is what you get"!  If the parameters in the neutrino sector
lead to some neutrino masses being in the eV range, then $\Omega_\nu$ is an
important component of $\Omega$.  If that is not the case, 
then $\Omega_{\nu}$ is not important.
So, from this view point, there is no difference in pedigree 
between dark matter which is a significant component for a range of particle physics
parameters (like the LSP), or relics which are important only 
for the specific value of some particle
physics parameters (like a KeV gravitino).

Adopting this point of view then, it is perfectly sensible to have various
particle physics excitations (say: baryons, neutralinos and neutrinos) play an
important role in the Universe now.  This is a welcome result, which is
reinforced by the power spectrum of density fluctuations in the Universe.
This spectrum also suggests that there is more than one component which
contributes to the energy density of the Universe.  Indeed, present data
on this spectrum seems to be best fit by having a variety of matter
components contributing.  For instance, recent work by Primack and
collaborators\cite{Primack} suggests that the power spectrum of density
fluctuations is optimally fit by having
\begin{equation}
\Omega_{\rm B} = 0.05;~~~\Omega_{\rm CDM} = 0.75;~~~\Omega_{\rm HDM} = 0.20~.
\end{equation}

These results are particularly interesting since the existence, or not, of
HDM provides a critical constraint (arising from cosmology) on the particle
physics which determines the neutrino mass matrix.\footnote{This information,
of course, is of relevance for experiments looking for neutrino
oscillations.}  If it were really possible to establish the need for
neutrino hot dark matter, for example through the influence it has  on the angular
spectrum of the CMBR,\cite{Burns} then this, along with the constraints 
imposed by neutrino oscillation experiments would do much
to fix the shape of the neutrino mass spectrum.  
As we discussed earlier, if one can establish both the
need for neutrino hot dark matter (which necessitates probably that
$\sum_i m_{\nu_i} \simeq (5-6~{\rm eV})$) 
and of neutrino oscillations with small
mass squared differences, then one is forced into a world of nearly
degenerate neutrino masses, with $m_{\nu_i} \simeq 1-2~{\rm eV}$.  Such
a result would provide a compelling argument for renewing the direct
searches in tritium beta decay 
for electron neutrino masses in the eV range. 

\section{The Sakharov Conditions for Baryogenesis}

In a classic paper, in 1967, Andrei Sakharov\cite{Sakharov} discussed the
conditions necessary to obtain dynamically an asymmetry between matter
and antimatter in the Universe.  Sakharov's conditions for obtaining this
asymmetry are three-fold:\cite{Sakharov}
\begin{description}
\item{(i)} The underlying physical theory must possess processes that
violate baryon number (B is not conserved).
\item{(ii)} The interactions which lead to B-violation, in addition
must violate C and CP.
\item{(iii)} To establish this asymmetry dynamically, furthermore, the 
B-violating processes must be out of equilibrium in the Universe.
\end{description}
Let me comment briefly on each of these points.  First, it is pretty clear
that if B is conserved then the total number of baryons minus anti-baryons
is a constant in time.  In this case, then the
difference $n_{\rm B}-n_{\rm\bar B}$
is a constant that is set by some initial boundary conditions.  Thus
$\eta$ is not generated dynamically, but is just a reflection of these 
initial boundary conditions and one is left to wonder why one has a value
$\eta\sim 10^{-10}$.

Similarly, it is also quite understandable why the second Sakharov
condition is needed.  If C and CP are good symmetries, one can transform
$n_{\rm B}$ into $n_{\rm\bar B}$ by one of these symmetry transformations.
Hence, even if B were to be violated, but if C or CP 
were to be good symmetries,
then one could never obtain a non-vanishing value for $\eta$.

The third Sakharov condition is slightly more subtle, but is also readily
understandable physically.  Roughly speaking, B-violating decays serve to
create a matter-antimatter asymmetry.  However, this asymmetry is destroyed
by inverse decays.  In thermal equilibrium, the rates for B-violating
decays and their inverses are the same, hence $n_{\rm B}-n_{\rm\bar B}=0$.

It is useful to demonstrate this last fact explicitly.  The rate of change of
$\Delta n_{\rm B} = n_{\rm B}-n_{\rm\bar B}$ as a result of B-violating
processes, if these processes are in equilibrium, is given by
the thermodynamic equation
\begin{equation}
\frac{d\Delta n_{\rm B}}{dt} = \gamma_{\not{\mbox{B}}} e^{-\mu/T}-\gamma_{\not{\mbox{B}}}
e^{\mu/T}~.
\end{equation}
Here $\gamma_{\not{\mbox{B}}}$ is the rate of B-violation per unit 
volume and $\mu$
is the chemical potential.  At high temperatures, 
one can expand the exponential factors and the above expression
reduces to
\begin{equation}
\frac{d\Delta n_{\rm B}}{dt} \simeq -\frac{2\mu}{T} \gamma_{\not{\mbox{B}}}~.
\end{equation}
However, in this temperature regime, one has simply that
\begin{equation}
\Delta n_{\rm B} = \frac{4}{\pi^2} \mu T^2~.
\end{equation}
Hence
\begin{equation}
\frac{d\Delta n_{\rm B}}{dt} \simeq -\frac{\pi^2}{2}
\left(\frac{\gamma_{\not{\mbox{B}}}}{T^3}\right) \Delta n_{\rm B} =
-\frac{\pi^2}{2} \Gamma_{\not{\mbox{B}}} \Delta n_{\rm B}~,
\end{equation}
where $\Gamma_{\not{\mbox{B}}}$ is just the rate for B-violation, since $V = T^{-3}$.
Thus, it follows from (82) that
\begin{equation}
\Delta n_{\rm B} = (\Delta n_{\rm B})_0 \exp\left[-\frac{\pi^2}{2}
\Gamma_{\not{\mbox{B}}}t\right]~.
\end{equation}
Eq. (83) tells one that, if B-violating processes are {\bf ever} in
equilibrium, then these processes serve to {\bf destroy} any pre-existing
asymmetry $(\Delta n_{\rm B})_0$.  This is a very nice result\cite{Weinberg1}
since it tells us that the value of $\eta$ one computes dynamically, as a
result of B-violating processes going out of equilibrium, is
{\bf independent} of any initial asymmetry $(\Delta n_{\rm B})_0$.  Hence,
the observed value of $\eta$ in the Universe now depends only on the B-violating (and C- and
CP-violating) dynamics---due to particle physics---and on the
cosmology which drives these processes out of equilibrium in the early
Universe.

I examine next cosmological circumstances (along with the relevant particle
physics) which can lead to baryogenesis.

\section{Baryogenesis at the GUT Scale: Issues and Challenges}

Grand Unified Theories (GUTs) were the first theories which explicitly
realized Sakharov's conditions for baryogenesis.\cite{Yoshimura}  These
theories naturally contain B-violating processes which also violate
C and CP.  An example is provided by $SU(5)$,\cite{su(5)} in which the
fermions of each generation are members of a 
$\bar 5\sim (d^c_{\rm L};e_{\rm L} \nu_{\rm L})$ and a
$10\sim (u_{\rm L}d_{\rm L}; u^c_{\rm L};d^c_{\rm L})$ 
representation\footnote{It is convenient to describe all states in terms of
how their left-handed components transform, using that
$\psi_{\rm R}\sim \psi_{\rm L}^c$.} and the ordinary Higgs doublet
$(\phi^+\phi^0)$ is augmented by a Higgs triplet $\chi$ into a field
$5_{\rm H}\sim (\chi;\phi^+\phi^0)$---with $\chi$ transforming under
$SU(3)\times SU(2)\times U(1)$ as $\chi\sim (3,1)_{-1/3}$.  In $SU(5)$, the
Higgs quintet $5_{\rm H}$ can couple to the fermions in two separate ways
$[5_{\rm H}~5~\overline{10}~\hbox{and}~5_{\rm H}~10~10]$, with the 
corresponding complex Yukawa couplings being sources for C and CP violation.
These couplings allow the triplet Higgs field $\chi$ to decay to both the
$d\nu$ (B = 1/3) and $\bar u\bar d$ (B = -2/3) final states.  Hence, in
$SU(5)$ baryon number is clearly not conserved.

Because one knows experimentally that baryon number is conserved to high
accuracy,\footnote{The PDG\cite{PDG} gives a bound for the B-violating
decay $p\to \pi^0 e^+$ of $ \tau(p\to \pi^0 e^+)>5 \times 10^{32}$
years.} one knows that a theory like $SU(5)$, where the $SU(3)$, $SU(2)$ and
$U(1)$ forces are unified, must break down to $SU(3)\times SU(2)\times U(1)$
at a very high scale: $M_X\sim 10^{15}-10^{16}~{\rm GeV}$.\cite{Langacker}
This {\bf unification scale} $M_X$ is quite near the Planck scale
$M_{\rm P}\sim 10^{19}~{\rm GeV}$.  We know that at temperatures near the
Planck scale, $T\sim M_{\rm P}$, the Universe is expanding very rapidly.
Thus it is not surprising that the C, CP and B-violating decays of GUTs
have rates which are slow with respect to the expansion rate of the Universe
at $T\sim M_X$.  That is
\begin{equation}
\frac{\dot R}{R} \sim \frac{T^2}{M_{\rm P}} > \Gamma_{\rm B-viol.} ~~~~~~
(T\sim M_X)~.
\end{equation}
Hence, the processes alluded above in GUTs also fulfill Sakharov's third
condition---that the relevant B-, C-, and CP-violating interactions be out of
equilibrium in the early period of expansion of the Universe after the Big
Bang.

These qualitative features, however, in practice do not lead to successful
simple scenarios for baryogenesis at the GUT scale.  Although it is possible
to obtain $\eta\sim 10^{-10}$ in some GUT models, these models have a number of
generic difficulties which are worth discussing here.
Again, it is useful to consider the $SU(5)$ example alluded above to help
focus on the source of these difficulties.

In $SU(5)$, the ratio $\eta$ is generated through the out of equilibrium decay
of the Higgs triplet $\chi$ at temperatures $T\sim M_X$.  One has
\begin{equation}
\eta = \frac{n_{\rm B}-n_{\bar{\rm B}}}{n_\gamma} \simeq A\Delta {\rm B}_\chi~.
\end{equation}
Here $A$ is a kinematical/dynamical factor related to the way the $\chi$
decays go out of equilibrium, while 
$\Delta B_\chi$ is the baryon asymmetry proper:
\begin{equation}
\Delta B_\chi = \sum_f
\frac{B_f\{\Gamma(\chi\to f)-\Gamma(\bar\chi\to\bar f)\}}
{\Gamma_\chi^{\rm total}}~,
\end{equation}
reflecting the differences in the weighted ratio of $\chi$ and $\bar\chi$
decays into particular final states $f$ with different baryon number 
$B_f$.\footnote{In the example discussed above $f=d\nu$ or $f=\bar d\bar u$.}
It should be clear from the form of Eq. (86) that $\Delta B_\chi$ vanishes if C
or CP is conserved, since 
then $\Gamma(\chi\to f)=\Gamma(\bar\chi\to\bar f)$.

It is easy to see that, for the example in question, one has simply
\begin{equation}
\Delta B_\chi = r-\bar r
\end{equation}
where
\begin{equation}
r = \frac{\Gamma(\chi\to d\nu)}{\Gamma_\chi^{\rm total}}~; ~~~~~~
\bar r = \frac{\Gamma(\bar\chi\to\bar d\bar \nu)}{\Gamma_\chi^{\rm total}}~.
\end{equation}
Eq. (87) has three characteristics:
\begin{description}
\item{(i)} It vanishes if there is no C or CP violation.  This is obvious,
since then $\bar r = r$.
\item{(ii)} $\Delta B_\chi$ vanishes also if one includes {\bf only} lowest
order processes.  Again this is easy to see since, at tree level,
$r=\bar r$.
\item{(iii)} Finally, and less obviously, $\Delta B_\chi$ also vanishes if
the underlying $\chi$-decays do not have an s-channel 
discontinuity.
\end{description}

\begin{figure}
\centerline{
\epsfig{file=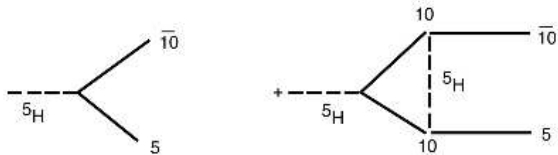,width=3.5in}
}
\caption[]{Graphs contributing to $r$ at one-loop order}
\end{figure}

One can see these three conditions at work by examining schematically the 
contribution to $r$ in the $SU(5)$ model we discussed earlier.  At one-loop
order, these contributions are given by the graphs shown in Fig. 15.  Let
us denote by $\gamma_0$ the rate associated with the tree graph decay in
Fig. 15 and by $\gamma_1I(M_\chi^2-i\epsilon)$ the contribution of the one-loop
graph.  In general both $\gamma_0$ and $\gamma_1$ are intrinsically complex as 
a result of the complex $\chi$ couplings, while the dynamical quantity
$I(M_\chi^2-i\epsilon)$ has an imaginary part as a result of the associated
loop integration.  A simple calculation, using Fig. 15, gives for the rate
difference $r-\bar r$ the expression:
\begin{eqnarray}
(r-\bar r) &\sim& |\gamma_0+\gamma_1 I(M^2_\chi-i\epsilon)|^2 -
|\gamma_0^*+\gamma_1^* I(M_\chi^2-i\epsilon)|^2 \nonumber \\
&\sim& {\rm Im}~\gamma_0\gamma_1^*~~{\rm Im}~
(I(M_\chi^2-i\epsilon))~.
\end{eqnarray}
One sees that this rate difference vanishes unless there is {\bf both}
an intrinsic CP violating phase difference in the couplings involved
[Im $\gamma_0\gamma_1^*$] as well as some imaginary part [Im $(I(M_\chi^2-i\epsilon))$
] arising from the (one-loop) scattering dynamics.  In view of Eq. (89), one
sees that the ratio $\eta=\Delta n_{\rm B}/n_\gamma$ is proportional to
\begin{equation}
\eta = A\Delta B_\chi =
A(r-\bar r)\sim A~{\rm Im}~\gamma_0\gamma_1^*~{\rm Im}~I~.
\end{equation}

The RHS of Eq. (90) embodies the essence of GUT baryogenesis.  The ratio
$\eta$ depends on the out of equilibrium dynamics [through $A$] and it
vanishes unless there is both an intrinsic CP and C violating phase
[Im $\gamma_0\gamma_1^*$] and the GUT dynamics is rich enough to generate
an s-channel discontinuity\break [Im $I(M^2_\chi-i\epsilon)$].  The knowledge of
each of these individual pieces is clearly model-dependent and quite 
rudimentary, since we have no direct evidence for the existence of any
GUTs~!  Thus, at this stage, it is really not possible to deduce a firm
prediction for $\eta$.  Even so, in general, one finds $\eta$ to be too small
unless one further complicates the GUT dynamics.

Let me illustrate the above point in the, by now familiar, $SU(5)$ context.  
Without loss of generality one can make the $5_{\rm H}~5~\overline{10}$
Higgs coupling matrix $f$ real.  Then it is easy to show that (for 3
families) the $5_{\rm H}~10~10$ coupling matrix $h$ has 3 
phases.\cite{GUTphases}  So GUTs, because they involve further Higgs
couplings, have more phases than the 3-family CKM phase connected with the
couplings of the Higgs doublet $\Phi$ to quarks.  Even so, in this model,
one cannot generate an intrinsic CP violating phase at one loop order.
The tree and one-loop level contributions in Fig. 15, corresponding to the
process $5_{\rm H}\to 10~\bar 5$, give
\begin{equation}
\gamma_0\sim f~; ~~ \gamma_1\sim fhh^{\dag}~.
\end{equation}
Hence
\begin{equation}
{\rm Im}~\gamma_0\gamma_1^*\sim {\rm Im}~{\rm Tr}~fhh^{\dag} f^{\dag} = 0~.
\end{equation}

One can check that other possible contributions to the decay
$5_{\rm H}\to 10~\bar 5$, involving gauge exchange in the $t$-channel rather
than Higgs exchange, are similarly relatively real.  As shown in Fig. 16, one
can eventually\cite{DG} obtain a non-vanishing $\eta$ for this model at
higher order, from the interference of a tree-level process with a
{\bf 3-loop} process.  The resulting $\Delta B_\chi$, however,
\begin{equation}
\Delta B_\chi\sim {\rm Im}~{\rm Tr}[h^{\dag} ff^{\dag} hff^{\dag} h^{\dag} h]
\end{equation}
has such a large number of Yukawa couplings that $\eta$ is at
best of $O(10^{-15})$.\cite{DG}\footnote{If one invokes a fourth generation
of quarks and leptons,\cite{ST} it is possible to boost up $\eta$ to the
desired $O(10^{-10})$ level even in this simple model.}

\begin{figure}
\centerline{
\epsfig{file=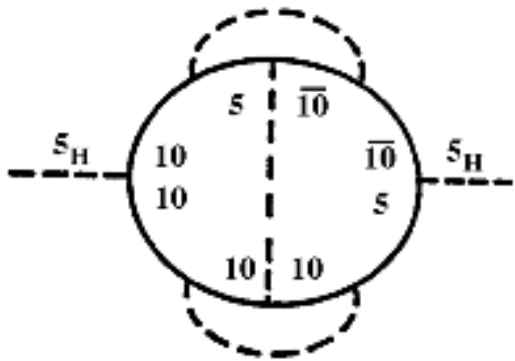,width=5in}
}
\caption[]{Interference graph giving $r \neq 0$ in $SU(5)$}
\end{figure}

This difficulty can be remedied by using more elaborate GUTs (or including
more low-energy states).  However, there are two further generic problems
connected with these types of models which serve to dampen the enthusiasm
for attributing baryogenesis in the Universe to some GUT processes.  The first
of these additional problems is related to monopoles.  In general GUTs lead
to an overproduction of monopoles in the early Universe, badly violating
one of the main features we know about the Universe now--namely that the
present Universe's energy density is near the critical density
$\rho\sim \rho_c$.\cite{Preskill}

't Hooft and Polyakov\cite{monopole} showed that monopoles always form when
a symmetry group breaks down to a subgroup containing a $U(1)$ factor--like the
standard model group.  Hence, if GUTs exist, one expects that in the very
early Universe at $T\sim M_X$, during the GUT phase transition, magnetic
monopoles are formed.  Generically, these GUT monopoles are superheavy, having
a mass of order 
$M_{\rm M}\sim\frac{M_X}{\alpha_{\rm G}}\sim 10^{17}~{\rm GeV}$.
During the GUT phase transition, domains of the broken phase of the GUT
group form which are of typical size $\xi \sim \frac{1}{T_c}\sim 
\frac{1}{M_X}$.  The superheavy GUT monopoles physically correspond to
topological knots between these domains and hence have a density
$n_{\rm M}\sim \xi^{-3}$.  This density is comparable to the photon
density at this stage of the Universe
\begin{equation}
n_{\rm M}(T_c)\sim \xi^{-3}\sim T_c^3\sim n_\gamma(T_c)~.
\end{equation}
However, such a large monopole density is extremely problematic, because of
the large mass of the GUT monopoles
.\cite{Preskill}  Indeed, from (94) one deduces that
\begin{equation}
\Omega_{\rm M}\big|_{\rm now} \sim M_{\rm M} n_{\gamma}\big|_{\rm now} \sim
10^{21}
\end{equation}
completely in contradiction with what we know!

Inflation provides a resolution of the monopole problem by inflating
exponentially the size
of the domains---essentially reducing the monopole density to
one per observable Universe.  However, to 
re-establish $\eta$ in such a scenario,
one has to reheat the Universe after the inflationary period to temperatures
$T_{\rm reheat} \sim 10^{14}-10^{15}~{\rm GeV}$, which is 
difficult to achieve.\cite{reheat}\footnote{At such temperature the number
density of monopoles produced after reheating is heavily suppressed by
a Boltzmann factor.}  Thus, the monopole problem, even if it is resolved by
inflation, argues against GUT baryogenesis.

There is another argument which also provides ammunition against the idea
that the baryon asymmetry in the Universe was produced at the GUT scale.  As
we will discuss shortly in more detail, it turns out that quantum effects
in the electroweak interactions can lead to the violation of total fermion
number (B+L---violation).\cite{'tHooft}  In the middle 1980's Kuzmin, Rubakov and
Shaposhnikov (KRS)\cite{KRS} argued that these (B+L)-violating processes,
which are extremely weak at $T=0$, could become strong enough at temperatures
near the electroweak phase transition, $T\sim M_W$, to go back into
equilibrium in the Universe.  The return of (B+L)-violating processes into
equilibrium in the Universe at $T\ll M_X$ serves to {\bf erase} any
(B+L)-asymmetry produced in the Universe at temperatures of the order of the
GUT scale, $T\sim M_X$.  
Hence, only a (B-L) asymmetry produced by GUTs survives
to low temperatures.

This consideration kills, for example, the baryon number asymmetry one
imagined was produced in the $SU(5)$ example discussed earlier.  It is
easy to check that for $\chi$-decays
\begin{equation}
\Delta B_\chi = \Delta L_\chi = r-\bar r~,
\end{equation}
so that
\begin{equation}
\Delta n_{\rm B-L} = \Delta n_{\rm B}-\Delta n_{\rm L} = 0~.
\end{equation}
Thus, as a result of the KRS mechanism, in this case no baryon asymmetry
survives at low temperatures, even if such an asymmetry 
were to be generated at the GUT scale
by the out of equilibrium decays of the Higgs triplet $\chi$.  Of course,
one can invent more elaborate GUTs scenarios in which at the GUT scale one
produces both a (B+L)- and a (B-L)-asymmetry, thereby bypassing this
conundrum.\cite{B-L}

\section{The KRS Mechanism and Baryogenesis at the Electroweak Scale}

In this section I want to discuss further the KRS mechanism\cite{KRS} 
because, besides erasing any previous (B+L)--asymmetry, it is possible that
through this mechanism one can actually produce the observed baryon 
asymmetry in the Universe during the electroweak phase transition.  This is an
exciting possibility, and one that has received considerable attention in
recent years.\cite{egbreview}  In the Standard Model, both baryon number, B,
and lepton number, L, are classical symmetries.  That is, they are symmetries
of the Standard Model Lagrangian:\footnote{If neutrinos are massless, then
the individual lepton numbers associated with electrons, muons and taus
$(L_e,L_\mu,L_\tau)$ are also SM Lagrangian symmetries.}
\begin{equation}
{\cal{L}}_{\rm SM}
\begin{array}{c}
\\ [-.1in]
\longrightarrow \\ [-.1in]
\mbox{\tiny L,B}
\end{array} 
{\cal{L}}_{\rm SM}~.
\end{equation}
However, because of the chiral nature of the electroweak interactions, at the
quantum level both the baryon number current, $J_{\rm B}^\mu$, and the 
lepton number current, $J_{\rm L}^\mu$, are not conserved.  Hence neither B,
nor L, remains a good symmetry at the quantum level, although their difference,
B-L, is still a conserved quantum number.

\begin{figure}
\centerline{
\epsfig{file=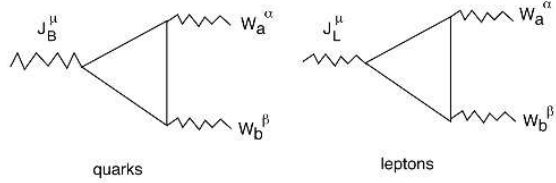,width=3.5in}
}
\caption[]{ Triangle graphs contributing to the B and L anomalies}
\end{figure}

The violation of B and L in the standard model comes about as a result of the
existence of chiral anomalies\cite{ABJ} in their respective currents.  For
our purposes, it suffices to focus only on the $SU(2)$ gauge field 
contribution to this anomaly.  The triangle graphs contributing to the anomalous
divergence of $J_{\rm B}^\mu$ and $J_{\rm L}^\mu$ are shown in Fig. 17 and
produce an equal divergence for both currents\cite{ABJ} 
\begin{equation}
\partial_\mu J_{\rm B}^\mu = \partial_\mu J_{\rm L}^\mu = -
N_g \frac{\alpha_2}{8\pi} W_a^{\mu\nu} \tilde W_{a\mu\nu}~,
\end{equation}
where $N_g$ is the number of generations and $\alpha_2 = g_2^2/4\pi$.
Clearly it follows then that
\begin{eqnarray}
\partial_\mu J_{\rm B-L}^\mu &=& 0 \nonumber \\
\partial_\mu J_{\rm B+L}^\mu &=& -N_g \frac{\alpha_2}{4\pi}
W_a^{\mu\nu} \tilde W_{a\mu\nu}~.
\end{eqnarray}

These equations, {\it per se}, do not 
automatically lead to a violation of (B+L)-number.
To get a change in B+L [$\Delta({\rm B+L}) \not= 0$] requires having
processes involving non Abelian gauge field configurations which have a non-trivial
index $\nu$:
\begin{equation}
\nu = \frac{\alpha_2}{8\pi} \int d^4x W_a^{\mu\nu}\tilde W_{a\mu\nu}~,
\end{equation}
since, in view of (100),
\begin{equation}
\Delta({\rm B+L}) = 2N_g~ \nu~.
\end{equation}
't Hooft\cite{'tHooft} was the first to estimate the size of the amplitudes which
contain gauge field configurations having such a non-trivial index $\nu$. These amplitudes arise in processes where the pure gauge field configurations
at $t=+\infty$ and $t=-\infty$ differ by a so-called, ``large" gauge
transformation\cite{CDG}.  In the $A^o = 0$ gauge, pure gauge fields can be
classified by how their associated gauge transformations go to unity at 
spatial infinity 
\begin{equation}
\Omega_n{(\vec r)}
\begin{array}{c}
\\ [-.1in]
\longrightarrow \\ [-.1in]
\mbox{\tiny ${\vec r}\to\infty$}
\end{array} 
e^{2\pi in}. \nonumber
\end{equation}  
One
can show that the index $\nu$ is related to the difference in the indices
$n_\pm$ characterizing the gauge vacuum configurations at $t=\pm\infty$
[$\nu = n_+-n_-$].\cite{CDG}  't Hooft's estimate\cite{'tHooft} of the size of
the amplitudes leading to (B+L) violation essentially involved the WKB
probability for tunneling from a vacuum characterized by index $n$ to one where
the index was $n+\nu$.  His result\cite{'tHooft}
\begin{equation}
A_{\rm (B+L)-violating} \sim {\rm exp}\left[-\frac{2\pi}{\alpha_2}\nu\right]
\end{equation}
has a typical WKB form, involving the inverse of the gauge coupling constant
squared in the exponent.  However, since the weak coupling constant
squared $\alpha_2$ is very small $(\alpha_2\sim 1/30)$, the result (104) is
extraordinarily tiny: $A_{\rm (B+L)-violating} \sim 10^{-80 \nu}$~!

't Hooft's result (104) is valid at $T=0$.  What Kuzmin, Rubakov and
Shaposhnikov\cite{KRS} realized was that the situation can be radically
different in the early Universe, when the (B+L)-violating processes 
happen in a non-zero thermal background.  When $T\not= 0$, the
gauge vacuum change needed for $\Delta({\rm B+L})\not= 0$ transitions to
happen can occur not only by tunneling, but also via a {\bf thermal
fluctuation}.  In this latter case, the transition probability is not given
by the square of the WKB amplitude (104), but instead by a Boltzman factor:
\begin{equation}
P_{\rm (B+L)-violation} \sim {\rm exp}\left[-\frac{V_o(T)}{T}\right]~.
\end{equation}
In the above, $V_o(T)$ is the (temperature dependent) height of the barrier
which separates inequivalent gauge vacuum configurations.

It turns out that one can estimate $V_o$ also by semiclassical methods;
in this case,
by using a static solution of the electroweak theory with minimum energy and winding number $n=1/2$.  This solution, first found by Klinkhamer and
Manton\cite{KM}, has been dubbed by them  a {\bf sphaleron}.  Essentially,
one takes $V_o(T)$ to be the energy associated with the sphaleron
configuration in the presence of a thermal bath:  $V_o(T) = E_{\rm sph}(T)$.
This energy has the typical form expected of a classical extended object.
It is proportional to the mass of the gauge field associated with the symmetry 
which suffers the breakdown and is inversely proportional to the gauge coupling
constant [c.f. the formula characterizing the monopole mass].  For the
sphaleron, one has
\begin{equation}
E_{\rm sph} = \frac{2M_W}{\alpha_2} f(M_{\rm H}/M_W)~,
\end{equation}
where $f$ is a function of order unity.

Because the $W$ mass, $M_W(T)$, vanishes as the temperature $T$ approaches
the temperature of the electroweak phase transition, $T\to T_{EW}$, the
probability of (B+L)-violating processes occurring in the Universe becomes
large as the Universe's temperature approaches $T_{EW}$.  This is basically
the fundamental observation made by Kuzmin, Rubakov and 
Shaposhnikov.\cite{KRS} That is, one expects that
\begin{equation}
P_{\rm (B+L)-violation}(T) \sim {\rm exp}\left[-\frac{E_{\rm sph}(T)}
{T}\right] 
\begin{array}{c}
\\ [-.1in]
\longrightarrow \\ [-.1in]
\mbox{\tiny $T \to T_{EW}$} 
\end{array}
1~.
\end{equation}

The original suggestion of KRS has been confirmed subsequently by much more
detailed calculations.\cite{Fratz}  Furthermore, one has found also a fast rate
for (B+L)-violation, above 
the temperature of the electroweak phase 
transition.\cite{Fabare}  These results are summarized below in a pair of
formulas giving the transition probability per unit volume, per unit time, for
temperatures below and above the temperature of the electroweak phase
transition.  One finds:
\begin{eqnarray}
\gamma_{\rm (B+L)-violation} &=& C\left[\frac{M_W^7}{(\alpha_2T)^3}\right]
{\rm exp}-\frac{E_{\rm sph}(T)}{T} ~~~ (T<T_{EW}) \nonumber \\
\gamma_{\rm (B+L)-violation} &=& C^\prime [\alpha_2T]^4 ~~~
(T>T_{EW})~,
\end{eqnarray}
where $C$ and $C^\prime$ are constants of order one.  These results imply that
the rate of (B+L)-violating processes, originating in the standard model, is
more rapid than the Universe's expansion rate $H\sim T^2/M_{\rm P}$ for rather
a large temperature interval:\cite{Rage}
\begin{equation}
\Gamma_{\rm (B+L)-violation} = \frac{\gamma_{\rm (B+L)-violation}}{T^3}
>H ~~ {\rm for}~~
T_{EW}\sim 10^2~{\rm GeV} \leq T \leq 10^{12}~{\rm GeV}~.
\end{equation}
A consequence of the above is that {\bf any} (B+L)-asymmetry established 
above $T_{\rm max} \sim\alpha_2^4M_{\rm P} \sim 10^{12}~{\rm GeV}$ 
(by, for example, some GUT processes) will get washed out.

Given these results, two possibilities emerge for trying to explain the observed
value of $\eta\sim 10^{-10}$:
\begin{description}
\item{i)} The baryon-antibaryon asymmetry underlying $\eta$ is the
result of a (B-L), or perhaps simply an L, asymmetry generated at high
temperatures.  Since the baryon number can be written as
\begin{equation}
B = \frac{1}{2} (B+L) + \frac{1}{2} (B-L)~,
\end{equation}
and all the (B+L)-asymmetry is erased by the KRS mechanism, one needs to
have some (B-L) asymmetry produced at high temperature to generate a 
non-vanishing value for $\eta$ now.
\item{ii)} The observed value of $\eta$ is the result of processes
occurring at the electroweak phase transition.  Baryogenesis 
is simply the reflection
of the violation of (B+L) in the standard model--{\bf electroweak 
baryogenesis}.
\end{description}

For the remainder of this section, I want to discuss this latter possibility.
This is a very intriguing suggestion\cite{Shap} and one which has generated
an enormous amount of interest recently.\cite{egbreview}  It is clear that
to be able to generate $\eta$ at the electroweak phase transition, one needs
this transition to be of {\bf first order}, so as to get a deviation from
thermal equilibrium.  As one goes through the phase transition, the Higgs VEV,
which vanished above $T_{EW}$, jumps to a non-zero value $\langle\phi(T^*)\rangle$ for temperatures below that of the electroweak
phase transition.

However,
to try to obtain through this non-equilibrium processes $\eta\sim 10^{-10}$
needs much more than just having a first-order phase transition.
There are actually two other main requirements.  First, one must make sure
that the asymmetry $\Delta n_{\rm B+L}$ created at the 
electroweak phase transition {\bf does not} get erased by having the
(B+L)-violating processes still be in equilibrium at $T^*$.  This requires
that
\begin{equation}
\Gamma_{\rm (B+L)-violation}(T^*) =
C\left[\frac{M_W^7}{\alpha_2^3T^{*6}}\right]
{\rm exp}\left[-\frac{E_{\rm sph}(T^*)}{T^*}\right]
< H(T^*) \sim \frac{T^{*2}}{M_{\rm P}}~.
\end{equation}
Numerically, this condition is equivalent to the requirement\cite{egbreview}
\begin{equation}
\frac{E_{sph}(T^*)}{T^*} \geq 45 ~~ \hbox{or} ~~
\frac{\langle\phi(T^*)\rangle}{T^*} \geq 1~.
\end{equation}
That is, to avoid erasure of the produced $\Delta n_{\rm B+L}$ after the
electroweak phase transition, this transition must be {\bf strongly first
order}, giving rise to a large jump for the Higgs VEV.

\begin{figure}
\centerline{
\epsfig{file=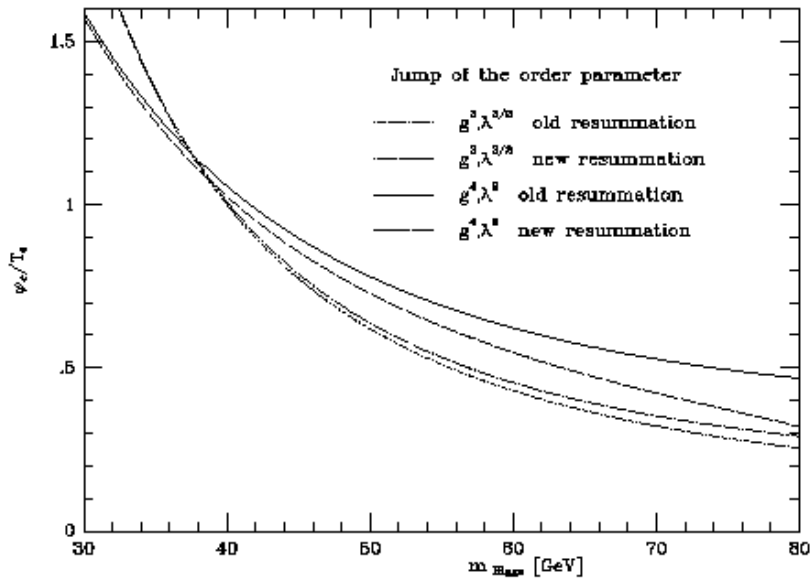,width=5in}
}
\caption[]{Jump in the order parameter in units of the critical temperature}
\end{figure}

One can compute the jump in $\langle\phi(T^*)\rangle$ from the 
temperature-dependent effective potential for the electroweak theory.
Although there are a number of uncertainties in this calculation,\cite{Buch}
it is now generally agreed\cite{egbreview} that in the Standard Model with
only one Higgs doublet one cannot get a sufficiently strong first-order
phase transition, unless the Higgs boson is light.  The present LEP bounds
on the Higgs boson mass, $M_H > 77.5~{\rm GeV}$\cite{Higgsbound} are already
sufficiently strong so as to rule out this simplest version of the Standard
Model as the source of the Universe's matter--anti-matter asymmetry.
This is made clear by Fig. 18, taken from\cite{Buch}, which shows that for
$M_H\sim 75~{\rm GeV}$ one expects $\langle\phi(T^*)\rangle/T^* \simeq
0.5$, in contradiction with the requirement of Eq. (112).  For such
high values of the
Higgs mass, standard model processes could create a baryon asymmetry
$\Delta n_{\rm B+L}$ at the electroweak phase transition, but this asymmetry
would
then get destroyed again at $T^*$, since the rate of (B+L)-violation is still
quite fast at this temperature compared to the Universe's expansion rate.

The situation, in this respect, is considerably better in supersymmetric
extensions of the Standard Model.  Carena, Quiros and Wagner, \cite{CQW}
for example, recently showed that the inequality (112) can be satisfied in
the MSSM, provided that one has a light stop, as well as small values for
tan $\beta$ and a reasonably light Higgs.  The values for $m_{\tilde t}$
and  $m_h$ required by \cite{CQW} to allow for baryogenesis at the weak
scale are low enough that these particles should be observable in the  near future already at
LEP 200 and/or 
at the Tevatron, when the Main Injector is put into operation. So perhaps one may, in this way, soon get indirect evidence for electroweak baryogenesis.

The second important requirement for believing that baryogenesis occurred at
the electroweak scale is to actually be able to carry out a detailed
dynamical calculation for $\eta$, yielding $\eta\sim 10^{-10}$.  Even if
the electroweak phase transition is sufficiently strongly first order, so
that the produced 
$\Delta n_{\rm B+L}$ is not erased, it is not obvious that one can produce a
big enough $\Delta n_{\rm B+L}$ so as to give $\eta\sim 10^{-10}$.

In the case of electroweak baryogenesis, the matter-antimatter asymmetry is
produced when bubbles of the true vacuum grow and fill up the Universe after
the electroweak phase transition.  As the Universe goes through the
(assumed) first-order electroweak phase transition, CP-violating processes
occurring in matter in the expanding bubble of true vacuum are crucial
to establishing a matter-antimatter asymmetry.  The rate of (B+L)-violation
per unit volume, $\gamma_{\rm (B+L)-violation}$, is rapid in the symmetric
vacuum surrounding the bubbles of true vacuum 
[$\gamma^{\rm outside}_{\rm (B+L)-violation}\sim (\alpha_2T)^4$].  However,
by assumption, if Eq. (112) holds, within the bubbles of true vacuum this
same rate is negligible [$\gamma^{\rm inside}_{\rm (B+L)-violation}\simeq 0$].
This rate difference is the key for establishing an asymmetry.

A full dynamical calculation of $\eta$ is very difficult.  Nevertheless, one 
can sketch pictorially what is going on simply by thinking of the expanding
true vacuum bubble as a wall sweeping through a plasma of quarks and 
antiquarks.  Because of CP-violating processes in the wall, the scattering
of quarks and antiquarks off the wall, as 
this wall sweeps through the plasma, is not the
same.  This difference in scattering will create an excess, say, of antiquarks
over quarks in the symmetric vacuum, $-\Delta n_q$, and an opposite excess of
quarks over antiquarks, $\Delta n_q$, in the true vacuum.  The fast
(B+L)-violating interactions in the symmetric vacuum, however, rapidly erase
the antiquark excess, $-\Delta n_q$, leaving a quark excess, $\Delta n_q$, in the true vacuum bubble.  This is the source of the baryon asymmetry.

There are many issues one has to resolve, or understand, to perform a
reliable calculation of the above processes.  For instance, what is the bubble
wall thickness?; what is the bubble value velocity?; etc.  Nevertheless,
although an actual calculation is difficult, one can at least arrive at an
order of magnitude estimate for $\eta$.  Recall that the asymmetry is
produced by the erasure of the antiquark excess in the symmetric vacuum.
Hence
\begin{equation}
\eta\sim\frac{\gamma_{\rm B+L~violation}}{T^3}\Big|^{T^*}_{\rm sym~vacuum} \sim
C^\prime\alpha_2^4T^*~.
\end{equation}
If one assumes, quite naturally, that all time scales in the problem are
set by $1/T^*$,\cite{HN} then the proportionality constant in 
Eq. (113)---besides the factor of $\alpha_2^4$---is set by the amount
of CP-violation in the quark and antiquark scattering off the bubble wall.
Calling this factor $\epsilon_{CP-violation}$, one arrives at the estimate
\begin{equation}
\eta\simeq[\alpha_2]^4\epsilon_{\rm CP-violation} \simeq
10^{-6}\epsilon_{\rm CP-violation}~.
\end{equation}
This estimate is confirmed by more detailed calculations, like those  
done recently by Huet and Nelson.\cite{HN}

Taking Eq. (114) as a reasonable guesstimate for electroweak baryogenesis,
one sees that this process is effective in creating a sufficiently large
baryon-antibaryon asymmetry only if there is enough CP-violation!
That is, one needs to generate in the bubble walls at least
$\epsilon_{\rm CP-violation}\sim 10^{-4}$.  It turns out, however,
that the standard model of flavor
violation---the CKM model\cite{flavor}---fails
miserably in this task.  In this case, because of the GIM mechanism,\cite{GIM}
there is no CP-violation unless the quark masses are different.  Whence, one
expects in this case that $\epsilon_{\rm CP-violation}$ contains a number of
GIM factors, which vanish if there is quark degeneracy.  In particular,
on general grounds, one expects
\begin{eqnarray}
[\epsilon_{\rm CP-violation}]^{\rm CKM} &\sim& 
\frac{[\lambda^6\sin\delta]}{[T^*]^{12}}\cdot
[(m_t^2-m_u^2)(m_t^2-m_c^2)(m_c^2-m_u^2) \nonumber \\
&\cdot&(m_b^2-m_d^2)(m_b^2-m_s^2)(m_s^2-m_d^2)]~.
\end{eqnarray}
The first square bracket contains the usual family mixing suppression 
factor,\footnote{In Eq. (115) $\lambda$ is the sine of the
Cabbibo angle and $\delta$ is the CP-violating phase in the CKM model.} \cite{Jarlskog} while the second factor involves a product of GIM
factors.  The result which follows from
Eq. (115), $\epsilon^{\rm CKM}_{\rm CP-violation}\sim 10^{-18}$,  is very small, falling far short of
what is needed. 

Eq. (115), however, may be too naive an estimate.  For example, one can avoid
altogether the GIM suppression factor of Eq. (115) in models where there are
some non-flavor violating sources of CP-violation at the electroweak scale.
Examples of such models are provided by multi-Higgs models, or models involving
a supersymmetric extension of the Standard Model.  Thus, if one believes that
baryogenesis is really an electroweak-scale phenomenon, one is forced to
contemplate theories which can provide a big enough $\epsilon_{\rm CP-violation}$.  Particle physics theories
which produce this result are necessarily enlargements
of the Standard Model, since the Standard Model itself cannot provide enough
$\epsilon_{\rm CP-violation}$.  If baryogenesis occurred at the
electroweak scale, these considerations argue that one is to expect both physics
beyond the Standard Model, and other CP-violating phases besides the 
standard CKM phase, already at this scale.  Conversely, if one were to find these
phenomena experimentally in the future, this would also provide indirect evidence
for electroweak baryogenesis.

\section{Generating a B-asymmetry from an L-asymmetry}

If the matter-antimatter asymmetry is not generated by electroweak baryogenesis
then, as I mentioned earlier, this asymmetry must arise from processes which
violate B-L at early times in the Universe.  Because all (B+L)-asymmetries
generated above temperatures of order $T\sim 10^{12}~{\rm GeV}$ are erased
by the KRS mechanism, purely L-violating processes effectively are equivalent
to (B-L)-violating processes.  However, these (B-L)-violating, or L-violating,
processes cannot themselves go back into equilibrium after the asymmetry is generated,
because that would serve again to erase this asymmetry.  As we shall see, this last
requirement has some (mild) implications for neutrino masses.

I want to illustrate this last point by discussing briefly a specific
model, due to Fukugita and Yanagida,\cite{FY} where what is violated is
actually lepton number.  In the Fukugita-Yanagida scenario there are
generic L-violating operators in the theory (arising from some GUT processes).
These operators give rise both to a neutrino mass for $\nu_{\rm L}$s and to
lepton number violating processes.  The simplest operator of this kind is
one which involves the usual $SU(2)\times U(1)$ left-handed lepton doublet
L for the first generation and the Higgs field $\Phi$:
\begin{equation}
{\cal{L}}_{\Delta L=2} = \frac{m_{\nu_e}}{v^2} 
L^TC\vec\tau L\cdot\Phi^{\dag}\vec\tau\Phi + {\rm h.c.}~.
\end{equation}
When $\Phi$ is replaced by its VEV, this term gives rise to a mass term for
the left-handed electron neutrino.  At the same time, this term also contributes
to the L-violating process $\nu_e\nu_e\to\Phi\Phi$.  A straightforward
calculation gives for the rate of L-violation
\begin{equation}
\Gamma_{\rm L-violation} =
\langle n\sigma(\nu_e\nu_e\to\Phi\Phi)\rangle \simeq
\frac{m^2_{\nu_e}}{\pi^3v^4} T^3~.
\end{equation}

If the masses of the neutrinos were to be large, this rate could be actually faster than
the Universe's expansion rate at temperatures below $T\sim 10^{12}~{\rm GeV}$,
where (B+L)-violating processes, due to the KRS mechanism\cite{KRS},
are themselves fast.  If this were to be the case, then {\bf no 
matter-antimatter asymmetry} would ever be generated at all!  The necessary
out of equilibrium condition $\Gamma_{\rm L-violation}< H\sim T^2/M_{\rm P}$, imposes
therefore a constraint on how large neutrino masses can be.  Using the
above result, one arrives at the bound\cite{RDPDallas}
\begin{equation}
m_{\nu_e} < \frac{0.4~{\rm eV}}{[T/10^{12}~{\rm GeV}]}~.
\end{equation}
That is, if neutrino masses are larger than this, then fast L-violating
processes (in conjunction with the KRS mechanism) can erase any previously
established matter-antimatter asymmetry.  Of course, this bound is really
rather soft in that it originates from only one possible type of
L-violating interaction.  Nevertheless, it is representative of a class of
generic bounds which exist if one does not attribute the observed 
matter-antimatter asymmetry to electroweak processes.\cite{RDPDallas}

This said, however, one should mention that it is possible to
avoid the above constraint.  The simplest way to do this\cite{Luty} is to
actually generate the matter-antimatter asymmetry of the Universe from
L-violating processes which go out of equilibrium much below
$T\sim 10^{12}~{\rm GeV}$.  This generally necessitates the introduction
of right-handed neutrinos, with the out of equilibrium decays of
$\nu_{\rm R}$ generating the required asymmetry.  The difficulty
in these scenarios, however, is producing a big enough asymmetry.\cite{Luty}  I will
not discuss this matter in detail here.  Suffice it to say that 
some successful models exist. These models have the peculiar feauture that
$\eta$ is driven by the CP-violating phases in the neutrino sector!  This last
fact is easy to understand since these phases, $\delta_\nu$. are the ones which
drive the lepton asymmetry [$\Delta n_{\rm L}\sim\sin\delta_\nu$] and through
the KRS mechanism, $\Delta n_{\rm B} = \Delta n_{\rm L}$.

\section{The Lessons of Baryogenesis for Particle Physics}

The above discussion of baryogenesis, either as a result of GUT models or through
Standard Model processes has been very speculative.  I believe, however, that
there are two overarching lessons one can draw from it.  The
first of these is that there are a plethora of particle physics scenarios
which can serve to generate a matter-antimatter asymmetry in the Universe.
Thus, Sakharov's intuition, that this asymmetry is dynamically generated and
not the result of some peculiar initial boundary condition, is most likely
true.

The second lesson one draws from these disquisitions is that what is crucial
for the whole issue of baryogenesis is the existence of {\bf other}
CP-violating phases, besides the usual CKM phase.  An important goal,
therefore, from a particle physics point of view is to try to discover these
phases experimentally.  This is a difficult, but perhaps not impossible task.
I would like to end these lectures by making a few remarks on this point,
particularly as it concerns electroweak baryogenesis.

Electroweak baryogenesis suggests the presence of {\bf flavor diagonal}
CP-violating phases.  These phasess arise quite naturally in multi-doublet
Higgs models and in supersymmetric extension of the Standard Model.
These phases, in general, do not contribute significantly to CP-violating
quantities which are sensitive to the CKM phase---like the CP-asymmetries in
B decays to CP self-conjugate states.\cite{BCP}  However, they can give
rather large contribution to some CP-violating parameters which are small
in the CKM model, like the electric dipole moment of the neutron.  For example,
supersymmetric extensions of the Standard Model give an electron dipole moment of the neutron which is of order\cite{edn}
\begin{equation}
d_n \simeq \frac{10^{-18}\sin\phi_{\rm SUSY}}{M^2_{\rm SUSY}({\rm GeV})}
{\rm ecm}~.
\end{equation}
In the above $\phi_{\rm SUSY}$ and $M_{\rm SUSY}$ are generic SUSY phases
and masses.  One sees that for $M_{\rm SUSY} \sim 100~{\rm GeV}$, the
present limits for the electric dipole moment of the neutron, of order 
$d_n \leq 10^{-25}~{\rm ecm},$\cite{PDG} requires $\phi_{\rm SUSY} \leq
10^{-3}$.  In fact, there is no real explanation why the SUSY violating
phases should be so small.\cite{susyphases}  So, it is obviously very important
that one should push the experimental limit for $d_n$ beyond $10^{-25}~{\rm ecm}$,
as a supersymmetric signal may just be lurking around the corner!
Unfortunately, this is a very difficult task in practice as experiments may
have already reached their ultimate sensitivity limit.

Fortunately, $d_n$ is not the only quantity which is sensitive to
flavor diagonal, CP-violating phases.  Another interesting
measurable quantity, which
perhaps is experimentally more accessible, is the transverse muon
polarization in the decays $K^+\to\pi^0\mu^+\nu_\mu$.  This quantity measures
the $T$-violating correlation:
\begin{equation}
\langle p_T^\mu\rangle = \langle\vec s_\mu\cdot(\vec p_\mu\times
\vec p_{\pi}\rangle~.
\end{equation}
Although $\langle p_T^\mu\rangle$ is not a purely CP-violating signal,
the final state interactions in this decay which could also produce a
transverse polarization are negligibly small
[$\langle p_T^\mu\rangle_{\rm FSI} < 10^{-6}$\cite{Z}].  Thus a
measurement of this quantity should test CP-violation.

What is interesting about $\langle p_T^\mu\rangle$ is that this quantity
vanishes in the Standard Model.\cite{Leuver}  However, if there are
CP-violating effective scalar interactions arising from physics beyond the
standard model, one can get a significantly large transverse muon 
polarization.  For instance, Grossman\cite{Grossman} finds that CP-violating
phases in the Higgs sector (in multi-Higgs models) which satisfy the present
bound on $d_n$, give a bound on $\langle p_T^\mu\rangle \leq 10^{-2}$.  
Remarkably, the present bounds on $\langle p_T^\mu\rangle$
are precisely at this level\cite{BNL}
\begin{equation}
\langle p_T^\mu\rangle = (-3.1\pm 5.3)\times 10^{-3}~.
\end{equation}
There is an experiment underway at KEK at the moment which hopes to push
the error on $\langle p_T^\mu\rangle$ to perhaps as low as
$\delta\langle p_T^\mu\rangle \sim 5\times 10^{-4}$.\cite{KEK}  Whether this can be
achieved remains to be seen.  However, in the near term, perhaps this is the
best chance for finding some non-CKM sources of CP-violation.  This is an
exciting and important discovery window.  If found, a non-vanishing value
for $\langle p_{\rm T}^\mu\rangle$ would have profound implications, not only
for particle physics but also for cosmology.

\section{Acknowledgments}

I am grateful to Ahmed Ali for having invited me to lecture at the
Nathiagali Summer College which proved to be a remarkable experience, particularly
because of the wonderful hospitality shown to me there.  This work is
supported in part by the Department of Energy under Grant No. 
DE-FG03-91ER40662, Task C.

\end{document}